\documentstyle[pra,aps,preprint]{revtex}
\tightenlines
\begin{document}
\draft
\title{Theory and experiment of the ESR of {\rm Co}$^{2+}$ in {\rm Zn}$_{2}${\rm %
(OH)PO}$_{4}$ and {\rm Mg}$_{2}{\rm (OH)AsO}_{4}$}
\author{M.E. Foglio, M.C dos Santos and G.E. Barberis}
\address{Instituto de F\'{i}sica ''Gleb Wataghin'', UNICAMP,\\
13083-970, Campinas, S\~{a}o Paulo, Brazil.}
\author{J.M. Rojo, J.L. Mesa, L. Lezama and T. Rojo}
\address{Departamento de Qu\'{i}mica Inorg\'{a}nica. Universidad del Pa\'{i}s\\
Vasco. \\
48080 Bilbao, Spain}
\date{\today }
\maketitle

\begin{abstract}
Experiments of Electron Spin Resonance (ESR) were performed on {\rm Co}$%
^{2+} $ substituting {\rm Zn}$^{2+}$ or {\rm Mg}$^{2+}$ in powder samples of
{\rm Zn}$_{2}${\rm (OH)PO}$_{4}$ and {\rm Mg}$_{2}{\rm (OH)AsO}_{4}$. These
two compounds are iso-structural and contain the {\rm Co}$^{2+}$ in two
environments of approximately octahedral and trigonal bipyramid structures.
The observed resonances are described with a theoretical model that
considers the departures from the two perfect structures. It is shown that
the resonance in the penta-coordinated complex is allowed, and the crystal
fields that would describe the resonance of the {\rm Co}$^{2+}$ in the two
environments are calculated. The small intensity of the resonance in the
penta-coordinated complex is explained assuming that this site is much less
populated than the octahedral one; this assumption was verified by a
molecular calculation of the energies of the two environments, with both
{\rm Co }and {\rm Zn} as central ions in {\rm Zn}$_{2}${\rm (OH)PO}$_{4}$.
\end{abstract}

\pacs{76.30.Fc,71.70.Ch,31.15.-p,31.15.Ct}

%\twocolumn

%\narrowtext

\section{\bf INTRODUCTION }

\label{secI}

Mineral solid state chemistry offers an important contribution to
materials science\cite{Riffel} in the search for systems with new
and useful physical properties. The phosphate and arsenate
minerals crystallize in various structures, sometimes containing
several non-equivalent sites for the metals. The minerals of the
olivine group, with the {\it ABX}O$_{4}$ formula, have been known
for a long time\cite{Richmond}, and the adamite family, with
formula [{\it M}$_{2}$(O/OH)({\it X}O$_{4}$)], belongs to this
group and takes its name from the natural compound\cite{Hawthorne,Hill} {\rm %
Zn}$_{2}${\rm (OH)AsO}$_{4}$. The compounds studied in this work belong to
this family, and the cations can occupy two sites with rather different
environments, one being octahedral and the other penta-coordinated, so that
rather different magnetic properties could be expected when magnetic cations
are employed. The recently synthesized compounds {\rm Zn}$_{2}${\rm (OH)PO}$%
_{4}$\cite{Harrison}, {\rm Co}$_{2}${\rm (OH)PO}$_{4}$\cite{Harrison}, {\rm %
Mg}$_{2}{\rm (OH)AsO}_{4}$\cite{Keller,KellerHZ}, as well as the
natural {\rm Co}$_{2}{\rm (OH)AsO}_{4}$\cite{Riffel}, are of the
adamite type and present these two type of sites. We have then
found interesting to study the properties of the {\rm Co}$^{2+}$
ions as impurities in the two non-magnetic compounds, as a first
step in the understanding of the properties of the concentrated
compounds.

Electron Spin Resonance (ESR) is a fruitful technique to obtain
local information of the environment of the magnetic ion, and the
{\rm Co}$^{2+}$ ion is particularly useful for this kind of study
because its {\bf g} value has a strong crystal field dependence in
these compounds. To analyze the ESR measurements it is necessary
to have information about the splitting of the energy levels with
both the crystal field and the electronic Coulomb repulsion, and
to obtain this information we employed optical diffuse reflectance
measurements\cite{Rojo00}.

The experimental ESR powder\ spectra of {\rm Co}$^{2+}$ impurities
in both {\rm Zn}$_{2}${\rm (OH)PO}$_{4}$ and {\rm Mg}$_{2}{\rm
(OH)AsO}_{4}$ present two different sets of lines, one very
intense, and the other just observable. The average of the g
factors of the intense spectra is 4.15 in the two compounds, a
value close to the 4.33 expected for {\rm Co}$^{2+}$ in moderately
distorted octahedral symmetry\cite{AbragamB}, and it seems
reasonable to assign these spectra to that environment and apply
the
approach already employed\cite{GoniFB} in the study of the ESR of \ {\rm Co}$%
^{2+}$ in {\rm NH}$_{4}${\rm NiPO}$_{4}${\rm .6H}$_{2}${\rm O }, where the
crystal fields that reproduce the observed spectra were obtained. As the
remaining lines are very weak one should analyze whether they belong to the
penta-coordinated symmetry, and in that case the possible reason for their
small intensity.

We are not aware of any theory describing the ESR of {\rm
Co}$^{2+}$ in the penta-coordinated environment, a distorted
trigonal bipyramid, and we first calculated the crystal fields of
the perfect trigonal bipyramid following the existing
literature\cite{BeltranP,Palacios}. To analyze the distorted
complex we derived the normal modes of the trigonal bipyramid with
respect to the reference complex, and then obtained the
Jahn-Teller contributions \cite{JahnT,Jahn} to the crystal field
acting on the\ {\rm Co}$^{2+}$, that is generated by these modes.

In this calculation we have introduced a procedure that uniquely defines the
orientation and size of the two reference complexes, so that the normal
modes that describe their deformation are free from irrelevant rotations and
expansions.

These results were then employed to calculate the theoretical ESR spectra.
We found that for the system parameters obtained from the optical spectra we
should expect that the ground doublet be $M_{J}=\pm 1/2$, corresponding to
an allowed spectrum. The rather small intensity, of this type of spectra
seems to indicate a preference of\ {\rm Co}$^{2+}$ for the octahedral sites
in the crystal structure, a conjecture that was advanced in a preliminary
report\cite{RojoMLBR96} on the ESR of impurities of this ion in {\rm Mg}$_{2}%
{\rm (OH)AsO}_{4}$, and that is confirmed in the present work. Employing a
molecular calculation we have also verified that the formation energies of
the two type of complexes, with both {\rm Co} and {\rm Zn} as the central
ions, are compatible with this hypothesis.

The description of the experiments, and experimental analysis of the data is
presented in Section \ref{secII}. The theory of the ESR of the\ {\rm Co}$%
^{2+}$ in a distorted trigonal bipyramid is presented in Section \ref{secIII}%
, together with the theoretical analysis of the two type of complexes in the
two compounds. A discussion of our results is presented in Section \ref
{secIV} together with our conclusions.

\section{\bf EXPERIMENTAL }

\label{secII}

\subsection{Synthesis and characterization of the materials.}

Compounds with {\rm Co}$^{2+}$ substituting {\rm Mg} and {\rm Zn} in {\rm Zn}%
$_{2}${\rm (OH)PO}$_{4}$ and {\rm Mg}$_{2}{\rm (OH)AsO}_{4}$ were prepared
by hydrothermal synthesis, starting from the ({\rm M, Co})$_{3}$({\rm XO}$%
_{4}):8${\rm H}$_{2}${\rm O (M = Zn, Mg)} vivianites, previously
prepared as reported elsewhere\cite{Rojo1}. Approximately 0.200~g.
of these precursors were disaggregated in {\it ca}. 75 mil. of
water and were placed in a poly(tetrafluoroethylene)-lined
stainless steel container (about three
quarters full) under autogenous pressure. The reaction was carried out at 180%
$^{\circ}$ C and maintained for one week. The resulting microcrystalline
products were filtered off and washed with ether and dried in air.

The results of the analysis of {\rm Mg, Zn, Co, P, and As} by inductively
coupled plasma atomic emission spectroscopy (ICP-AES) are in good agreement
with the proposed formulae. The compounds were also characterized by X-ray
powder diffraction, using the Rietveld method. The diffractograms were
indexed with the $P_{nnm}$ space group and the lattice parameters $%
a=8.042(3) $ \AA , $b=8.369(2)$ \AA\ and $c=5.940(2)$ \AA\ for the
phosphate compound and $a=8.286(2)$ \AA , $b=8.594(2)$ \AA\ and
$c=6.051(1)$ \AA\ for the arsenate. The parameters of the
phosphate compound are only slightly different from those given in
reference~\onlinecite{Harrison}, while those of the arsenate
coincide with the data published by other authors~\cite
{Keller,KellerHZ}; the results obtained in the last three
references were performed in single crystals. The X-ray powder
pattern was recorded employing Cu Ka radiation with a PHILIPS
X'PERT automatic diffractometer, with steps of $0.02{{}^{\circ }}$
in 2$\Theta $ and fixed-time counting of 1 s in the $5<2\Theta
<70{{}^{\circ }}$ range. We preferred to use our own experimental
parameters in the present paper.

\subsection{Optical studies.}

The necessary optical data were obtained from diffuse reflectance
experiments, performed in a CARY 2415 UV-VIS-IR spectrometer,
controlled with a VARIAN DS15 workstation, in the 5000 -- 50000\
{\rm cm}$^{-1}$ wavenumber region\cite{Rojo00,RojoMPLAR97}. The
whole of the optical data used in this work was recorded at room
temperatures, and all the relevant data that was necessary in the
present work is in the tables \ref{T1p} and \ref{T6}. Figure
~\ref{Fig001} presents the experimental data for both the
phosphate and the arsenate Co compounds studied in this paper. The
system parameters of the octahedral complexes are slightly
different from those already published\cite{Rojo00,RojoMPLAR97},
because they were obtained from the optical spectra after
including a spin orbit correction in the ground orbital
level\cite{Note1}.

\subsection{Electron Spin Resonance (ESR).}

\label{secIIC}

The ESR spectra were performed at {\it X} Band on a Bruker ESP300
spectrometer. Cooling and temperature control of the samples were obtained
with a standard OXFORD helium continuous-flow cryostat, included in the
microwave cavity. Magnetic field measurements were done simultaneously with
the ESR spectra recording, using a Bruker ER035M NMR gaussmeter. The
resonant frequency of the cavity was measured with a Hewlett-Packard 5352B
microwave frequency counter.

Only powder spectra could be measured for the two systems studied here
because it was not possible to obtain single crystals, and small
concentrations of Co (1\% in the arsenate and 0.1\% in the phosphate)
substitute the metals in the two lattices. The curves denoted with (a) in
Figs.~\ref{Fig01} and \ref{Fig02} show the measured ESR spectra for the two
samples, recorded at 4.2 K, and they both clearly show three sets of lines
with a well defined hyperfine structure that identifies the {\rm Co}$^{2+}$
ion. There are also some extra lines, rather weak in the phosphate but more
intense in the arsenate, that preclude an automatic fitting of the spectra.
We have then simulated the powder spectra of the hexa-coordinated {\rm Co}$%
^{2+}$ with a program which allows any symmetry, line positions, hyperfine
tensor, and linewidth anisotropies, and our best results, plotted in the
curves (c) of Figs.~\ref{Fig01} and \ref{Fig02}, correspond to the g-values
shown in rows a) of table~\ref{T01} Their values and positions are also
shown by arrows below the simulated curves (c).

The extra lines near 200 mT in the phosphate show an hyperfine
structure typical of the {\rm Co}$^{2+}$, and are given in more
detail in the inset of Fig.~\ref{Fig01}. The remaining lines in
the two compounds are rather broad and show a collapsed hyperfine
structure. The curves (b) in the Figs.~\ref {Fig01} and
\ref{Fig02} show the sum of the simulated spectra of the
hexa-coordinated {\rm Co}$^{2+}$ in (c), plus a simulation of the
penta-coordinated {\rm Co}$^{2+}$that employs the $g$- values
given in rows b) of table \ref{T01} and is adequately renormalized
to account for the smaller relative concentration of the last
compound. These $g$ values have rather large errors, and their
positions are shown by arrows above the measured spectra (a) of
Figs.~\ref{Fig01} and \ref{Fig02}. In the inset of Fig.
\ref{Fig01} it is also shown the detail of the hyperfine structure
near 200 mT both in the experimental and in the simulated
spectrum.

The assignment of the extra lines to the penta-coordinated complex shall be
further discussed in section \ref{B5}.

\section{\bf THEORETICAL DISCUSSION}

\label{secIII}

\subsection{Hexa-coordinated Co}

\label{ssec:A}We shall first discuss the hexa-coordinated ${\rm Co}^{2+}$
ions, that are surrounded by six oxygens in a fairly regular octahedron with
positions given in table \ref{T1}. In the present case we have a powder
spectra, and we could only measure the three principal values $g_{i}$ of the
${\bf g}$ tensor (see table \ref{T01}). As in a previous work \cite{GoniFB}
we shall consider the effect that the crystal field generated by the normal
modes of the octahedron has on the gyromagnetic tensor ${\bf g}$. This
method systematizes the procedure, and in table~\ref{T2} we give the normal
modes that reproduce the experimental values of the three $g_{i}$. We shall
then choose a reference perfect octahedron centered in the ${\rm Co}^{2+}$,
calculate the normal modes corresponding\ to the crystallographic positions
of the O, and compare them in table \ref{T2} with those obtained from the
experimental spectra.

Only the normal coordinates that are invariant against inversion with
respect to the center of the octahedron are necessary in the present
problem, \cite{VanVleck,GoniFB} and these are separated into the three sets $%
\{Q_{1}\},\{Q_{2},Q_{3}\}$ and $\{Q_{4},Q_{5},Q_{6}\}$, with the
corresponding $Q_{j}$ transforming respectively like the basis of the
irreducible representations $A_{1}$, $E$ and $T_{2}$ of the cubic group, as
given in table II of reference \onlinecite{Koster}.

The $^{4}F$ ground state of isolated ${\rm Co}^{2+}$ ($3d^{7}$) in a purely
octahedral crystal field splits into two orbital triplets $^{4}T_{1}$,$%
^{4}T_{2}$ and one orbital singlet $^{4}A_{2}$. Spin-orbit effects partially
lift the degeneracy of the $^{4}T_{1}$ triplet into one $\Gamma _{6}$, two $%
\Gamma _{8}$ and one $\Gamma _{7}$ subspaces, and the resonance
for the lowest doublet ($\Gamma _{6}$) is isotropic with
$g=4.33$\cite{AbragamB}. The addition of lower symmetry crystal
fields produce further splitting of the $^{4}T_{1}$ triplet,
giving six Kramer's doublets, and in most cases it is found that
the trace of the g tensor is close to the cubic isotropic
value;\cite{Tinkham(1)} in the present case the average g is 4.1537 for {\rm %
Co:Zn}$_{2}{\rm (OH)PO}_{4}$ and 4.153 for {\rm Co:Mg}$_{2}{\rm (OH)AsO}_{4}$%
. To understand these values it is sufficient to consider the ${\rm Co}^{2+}$
in pure octahedral symmetry, because the crystal fields of lower symmetry do
not change this value in our approximation. The calculation follows the same
lines given in reference\cite{GoniFB} and shall not be repeated here.

In the lowest order one obtains ${\bf g}$ from the matrix elements of the
Zeeman term in the $\Gamma _{6}$ subspace of the $^{4}T_{1}$ ground triplet.
The matrix elements of the orbital angular momentum ${\bf L}$ within a $%
T_{1} $ subspace are proportional to those of a $P$ term, but one should
note that the excited term $^{4}P$ is also of the $^{4}T_{1}$ symmetry, and
is mixed by the cubic field with the $^{4}T_{1}$ of the ground $^{4}F$ term.
If we indicate two states of $^{4}F$ and $^{4}P$ with $\phi _{i}$ and $\phi
_{i}^{\prime }$ respectively, such that they transform in the same way under
the cubic group, the states of the ground $^{4}T_{1}$ will be of the form $%
a\phi _{i}+b\phi _{i}^{\prime }$ . The values of the constants $a$ and $b$
can be obtained\cite{PeixotoF,Tucker} from the Racah parameter $B$ and the
crystal field parameter $D_{q}$, that were estimated\cite{Note1} from the
spectroscopic data and are given in table \ref{T1p}. With these values one
obtains $a=-0.9820$ and $b=0.1886$ for the phosphate, and the
proportionality constant of the angular momentum is then
\begin{equation}
\alpha =-1.5\ a^{2}+b^{2}=-1.4110.  \label{Et1}
\end{equation}

To analyze further the experimental ${\bf g}$ tensor, one could try and find
crystal field values that would reproduce the measured results, and a study
of this type was presented by Abragam and Pryce for the Cobalt Tutton salts.
\cite{AbragamP51} To simplify the study we present a model that describes
all the crystal fields acting on the ${\rm Co}\,$ as originating in the
crystal field of the six nearest ${\rm O}$ located at the vertices of a
deformed octahedron, obtained by displacement of the vertices of the
reference octahedron. If one neglects the mixing of other configurations
into the ground configuration $(3d)^{7}$, it is sufficient to keep only the
part of the crystal field $V$ that is even against inversion. We could then
write $V=\sum_{i=1}^{7}V({\bf r}_{i})$ , where $V({\bf r})$ would be the sum
of homogeneous polynomials of second and fourth order in the components $x,\
y,\ z$ of the electronic coordinates ${\bf r}$. Within our model, one could
then write\cite{VanVleck}
\begin{equation}
V({\bf r})=\sum_{j}Q_{j}\ V_{j}({\bf r})  \label{Et3}
\end{equation}
where the $Q_{j}$ and$\ V_{j}({\bf r})$ transform like the same
partners of irreducible representations of the octahedral
group\cite{Koster}. As the$\ V_{j}({\bf r})$ must be even against
inversion, the $Q_{j}$ must have the same property, an only the
six $Q_{j}$ with $j=1,6$ discussed at the beginning of this
section would appear in Eq. (\ref{Et3}), but we shall not consider
the identical representation $A_{1}$ because it does not modify
the
g tensor. The useful $V_{j}({\bf r})$ are given in Eqs.~(4-6) from reference %
\onlinecite{GoniFB}.

To study the effect that the $V({\bf r})$ given in Eq.(\ref{Et3})
has on the g tensor of, we shall employ second order perturbation
theory\cite{Pryce},
using both $V({\bf r})$ and the Zeeman term $H_{Z}=(g_{e}\ {\bf S}+{\bf L}).%
{\bf H} $ as perturbation. The change ${\bf \delta g}$ in the g tensor is
then obtained from
\begin{eqnarray}
{\bf S.\delta g.H} &=&\frac{2}{3}\ (g_{e}+\alpha )\frac{\mu _{B}}{\Delta }%
\left\{ -C_{E}\left[ \sqrt{3}Q_{2}\left( S_{x}H_{x}-S_{y}H_{y}\right)
+\right. \right.  \nonumber \\
&&\left. Q_{3}\left( 3S_{z}H_{z}-{\bf S.H}\right) \right] +C_{T}\left[
Q_{4}\left( S_{z}H_{y}+S_{y}H_{z}\right) +\right.  \nonumber \\
&&\left. \left. Q_{5}\left( S_{x}H_{z}+S_{z}H_{x}\right) +Q_{6}\left(
S_{x}H_{y}+S_{y}H_{x}\right) \right] \right\} ,  \label{Et7}
\end{eqnarray}
where $\mu _{B}$ is the Bohr magneton, $\Delta $ is the splitting between
the $\Gamma _{6}$ doublet and the lowest $\Gamma _{8}$ quadruplet in the
octahedral symmetry, and $\Delta =283$ cm$^{-1}$ in the {\rm P} compound.

The values of $C_{E}$ and $C_{T}$ are obtained by the same procedure
employed in reference\onlinecite{GoniFB}, obtaining $\langle r^{4}\rangle$
from the cubic field parameter $D_{q}$ and $\langle r^{2}\rangle $ from the
ratio $\langle r^{2}\rangle /\sqrt{\langle r^{4}\rangle }= 0.6544$ of the
calculated values\cite{Tucker}. For the {\rm Co-O }distance $R$ we used $%
R=2.11176$ \AA, corresponding to the reference octahedron defined below, and
we found the values $C_{E}=6436\ {\rm cm}^{-1}/$\AA \ and $C_{T}=-3666\ {\rm %
cm}^{-1}/$\AA . We can now calculate the crystal fields that would describe
the experimental values of ${\bf g}$ or, what is equivalent, the
corresponding normal modes within the approximations just discussed. As
there are more normal modes than data, we fix the relations $%
Q_{4}=Q_{3}=0.3044 Q_{6}$, which correspond to the normal modes calculated
below from the crystallographic positions, and we obtain a perfect fit to
the experimental values employing the normal modes given table \ref{T2}.

>From table \ref{T1p} we obtain the coefficients $a=-0.9824$, $b=0.1867$, and
$\alpha =-1.4128$ for {\rm Co}$_{2}{\rm (OH)AsO}_{4}$. The ESR data was then
adjusted with the normal modes coordinates given in table \ref{T2}, where we
used $R=2.1224\ \AA$, $Q_{4}=Q_{3}=0.4037Q_{6}$,\ $C_{E}=6287\ {\rm cm}%
^{-1}/ $\ \AA , $C_{T}=-3558\ {\rm cm}^{-1}/$ \AA \ and\ $\Delta=282\ {\rm cm%
}^{-1}$, for this compound.

To calculate the crystallographic normal modes of the octahedron it is
necessary to chose a reference perfect octahedron centered in the ${\rm Co}%
^{2+}$. To this purpose we consider the three normal modes of pure rotation,
\cite{VanVleck} $Q_{19}$, $Q_{20}$, and $Q_{21}$, and we chose the axes of
the reference octahedron so that these three normal coordinates are zero,
because they should not have any effect on the properties of the complex.
The value $R=2.11176$ \AA \ of the {\rm Co-O} distance in the reference
octahedron was chosen so that $Q_{1}=0$, and by this whole procedure we
obtain a unique reference octahedron and minimize the effect of irrelevant
rotations and expansions on the values of the normal modes. The direction
cosines of the three {\rm Co-O} directions in the reference octahedron are
given in table \ref{T3}, and the normal modes derived from the
crystallographic ionic positions given in table \ref{T1} are shown in the
third line of table \ref{T2}. The normal modes calculated from the
crystallographic position of the ${\rm O}$ in the octahedral complex are
different than those obtained from the experimental ${\bf g}$ tensor, given
in the first line of the same table. This result indicates that although the
nearest ${\rm O}$ to the ${\rm Co}$ are the main source of the cubic field,
\cite{VanVleck} the remaining non-cubic perturbations have strong
contributions due to the rest of the crystal. We conclude that the
experimental ${\bf g}$ tensor could be explained by the crystal field $V(%
{\bf r})$ of Eq. (\ref{Et3}) given in the axes of the reference octahedron
defined in table \ref{T3} with the $Q_{j}$ given in the row a) of table \ref
{T2}. The agreement is perfect because there are more free normal
coordinates than available ${\bf \delta g}$ components, but the theory
presented can only be considered a first approximation. In particular,
although the crystal field theory of point charges gives the right symmetry
properties, it is only a very rough description of the physics of the
problem.

Although we have not analyzed the hyperfine tensor in detail, we have
verified that its components are compatible with the normal modes necessary
to describe the ${\bf \delta g}$ tensor.

In the present calculation we have neglected the effect of the $^{4}T_{2}$
triplet, that contributes to $\delta {\bf g}$ in third order perturbation
(our calculation would be of the second order). This effect was calculated
by Tucker\cite{Tucker} who obtained contributions that are about 6\% of the
second order contribution for the $T_{2}$ deformation and about 13\% for the
$E$ deformation, and would therefore not alter substantially our conclusions.

\subsection{Penta-coordinated Co\label{B}}

\subsubsection{The crystal field of the trigonal bipyramid.\label{B1}}

The structure of the penta-coordinated complex of {\rm Co}$^{2+}$ is very
close to a trigonal bipyramid, and the positions of the {\rm \ Co} and of
the five {\rm O} are given with respect to the crystal axes in table \ref
{T4p}. Following a method similar to that employed in the octahedral case we
chose an orthogonal system of axes X,Y,Z, such that the normal modes
obtained from the crystallographic positions would not have contributions of
irrelevant rotations and expansions. The direction cosines of the axes of
this system with respect to the crystal axes {\bf a}, {\bf b} and {\bf c }%
are given in table \ref{T5}, and the coordinates of the six atoms in the
reference perfect trigonal bipyramid are given in table \ref{T4}. There are
two different {\rm Co-O }distances in the reference complex: $R_{a}$
corresponds to the three ligands in the XY plane (equatorial {\rm O}) and $%
R_{c}$ to the two along the Z axis (axial {\rm O}); their values for the
phosphate and arsenate are given in the caption of table \ref{T4}. Two
crystal field parameters $D_{s}$ and $D_{t}$ are necessary in the trigonal
bipyramid, and are given in the point charge model by:\cite{BeltranP,Wood}
\begin{eqnarray}
D_{s} &=&\frac{e}{14}\left[ \frac{4q_{c}}{R_{c}^{3}}-\frac{3q_{a}}{R_{a}^{3}}%
\right] \langle r^{2}\rangle ,  \nonumber \\
D_{t} &=&\frac{e}{168}\left[ \frac{16q_{c}}{R_{c}^{5}}+\frac{9q_{a}}{%
R_{a}^{5}}\right] \langle r^{4}\rangle ,  \label{E2.1}
\end{eqnarray}
where we shall use $q_{a}=q_{c}=-2e$. The crystal field potential $V_{cf}$
can be expressed by the usual formula
\begin{equation}
V_{cf}({\bf r})=\sum\limits_{kq}\sqrt{\frac{4\pi }{2k+1}}\sum\limits_{\ell
}q_{\ell }\frac{r_{\ell <}^{k}}{r_{\ell >}^{k+1}}Y_{kq}^{\ast }(\theta
_{\ell },\varphi _{\ell })\ C_{q}^{(k)}(\theta ,\varphi ),  \label{E2.2}
\end{equation}
where $Y_{kq}(\theta _{\ell },\varphi _{\ell })$ are the spherical harmonics
at the position of the $\ell $-th ligand and the $C_{q}^{(k)}(\theta
,\varphi )=\sqrt{4\pi /(2k+1)}Y_{kq}(\theta ,\varphi )$ are usually called
the Racah's rationalized spherical harmonics. In our actual calculation we
have employed the real combinations $C_{lm}(\theta ,\varphi )$ and $%
S_{lm}(\theta ,\varphi )$ that are proportional to $cos(m,\varphi )$ and $%
sen(m,\varphi )$ respectively\cite{Griffith}. In the absence of
the spin-orbit interactions one employs the irreducible
representations $\Gamma $ of the trigonal bipyramid to classify
the eigenstates $\left| \alpha ,S,L,\Gamma ,\gamma ,a\right\rangle
$ of the Hamiltonian, which are simply
related to the states $\left| \alpha ,S,L,M_{L}\right\rangle $ (the index $%
\alpha $ identifies the particular states with the same $S,L$). In table I
of reference~\onlinecite{BeltranP} we find that the irreducible
representations $A_{2}^{\prime }$, $A_{1}^{\prime \prime }$, $A_{2}^{\prime
\prime }$, $E^{\prime }$ and $E^{\prime \prime }$ are contained in the two
terms $^{4}F$ and $^{4}P$, and that the $\left| \alpha
,S,L,M_{L}\right\rangle $ states that generate the corresponding subspaces
are \{$\left| 3,3/2,3,0\right\rangle ,\left| 3,3/2,1,0\right\rangle $\}$%
\rightarrow A_{2}^{\prime }$, \{$\left| 3,3/2,3,\pm 3\right\rangle $\}$%
\rightarrow $ ($A_{1}^{\prime \prime }$, $A_{2}^{\prime \prime }$), \{$%
\left| 3,3/2,3,\pm 2\right\rangle $\}$\rightarrow E^{\prime }$ and \{$\left|
3,3/2,3,\pm 1\right\rangle ,\left| 3,3/2,1,\pm 1\right\rangle $\}$%
\rightarrow E^{\prime \prime }$. The Hamiltonian without spin
orbit interaction is diagonal in the partners $\gamma $ of each
irreducible representation $\Gamma $ and in the spin component
$M_{S}$, so it is not necessary to write them explicitly here. The
only $C_{q}^{(k)}(\theta ,\varphi )$ that contribute to
Eq.(\ref{E2.2}) in the perfect trigonal bipyramid have $k=0,2,4$
and $q=0$. To calculate the matrix elements of the Hamiltonian
that contains $V_{CF}=\sum_{i=1,7}V_{cf}({\bf r}_{i}),$ we have
used the standard tensorial operator techniques\cite{FanoR} as
well as the unitary operators obtained form Nielsen and Koster's
tables\cite{FanoR,NielsenK}, and we have verified that our matrix
coincides with that given in table II or reference
\onlinecite{BeltranP}.

Our main objective here is to find the gyromagnetic factors that one would
expect to measure in the penta-coordinated ${\rm Co}^{2+}$, and we shall
employ the spectroscopic data measured by diffuse reflectance to estimate
the parameters $B$, $D_{s}$ and $D_{t}$ for both {\rm Co:Zn}$_{2}{\rm (OH)PO}%
_{4}$ and {\rm Co:Mg}$_{2}{\rm (OH)AsO}_{4}$. In the two rows labeled a) of
table \ref{T6} we give the corresponding assignments of the transitions from
the ground $^{4}A_{2}^{\prime }$ to the levels with symmetry $%
^{4}A_{1}^{\prime \prime }$, $^{4}A_{2}^{\prime \prime }$, $^{4}E^{\prime
\prime }$, $^{4}E^{\prime }$, $^{4}A_{2}^{\prime }(P)$ and $^{4}E^{\prime
\prime }(P)$, where we use $(P)$ to indicate the higher levels of the same
symmetry.

>From the eigenvalues of the Hamiltonian in the absence of the spin-orbit
interaction, we find by trial and error the values of $B$, $D_{s}$ and $%
D_{t} $ that minimize the mean square deviation $\chi $ for the two systems,
and we give them in row b) of table \ref{T6p}. The transitions calculated
with these two sets of values are given in the two rows of table \ref{T6}
that are labeled b). The fitting is rather poor, and in particular the
transitions to the levels $^{4}A_{1}^{\prime \prime }$, $^{4}A_{2}^{\prime
\prime }$ and $^{4}E^{\prime \prime }$ fall below the range of the measuring
equipment. As an alternative we have fitted only the three highest
transitions, obtaining the values given in row c) of table \ref{T6p}, and
the corresponding values calculated with these two sets of parameters are
given in the two rows of table \ref{T6} that are labeled c). In the
following section we shall consider these two sets of values to estimate the
gyromagnetic factors for each of the two compounds.

\subsubsection{The spin-orbit interaction in the trigonal bipyramid.}

\label{B2}It is now essential to include the spin orbit interaction into the
calculation. The basis of the irreducible representations $\Gamma _{7}$, $%
\Gamma _{8}$ and $\Gamma _{9}$, of the double group $D_{3h}^{\ast }$ have a
simple expression in our system:\cite{Palacios} they are given by $\left|
d^{7}\alpha SLJM_{J}\right\rangle $, and in particular we have $\Gamma
_{7}(a)\equiv \left\{ \left| d^{7}\alpha SLJ\pm 1/2\right\rangle \right\} $,
$\Gamma _{7}(b)\equiv \left\{ \left| d^{7}\alpha SLJ\pm 11/2\right\rangle
\right\} $, $\Gamma _{8}(a)\equiv \left\{ \left| d^{7}\alpha SLJ\pm
5/2\right\rangle \right\} $, $\Gamma _{8}(b)\equiv \left\{ \left|
d^{7}\alpha SLJ\pm 7/2\right\rangle \right\} $, $\Gamma _{9}(a)\equiv
\left\{ \left| d^{7}\alpha SLJ\pm 3/2\right\rangle \right\} $ and $\Gamma
_{9}(b)\equiv \left\{ \left| d^{7}\alpha SLJ\pm 9/2\right\rangle \right\} $.
These states are easily obtained from the $\left| d^{7},\alpha
,S,M_{S},L,M_{L}\right\rangle $ calculated above by employing the 3-j or the
Clebsch Gordan coefficients. In the absence of magnetic fields the two
states of each Kramer's doublet have the same energy, and to calculate the
energies of the system it is enough to consider only the states with
positive $M_{J}$. As only the mixture of the $^{4}F$ and $^{4}P$ states is
important in our problem we shall consider only that subspace, and the
corresponding matrix of the total Hamiltonian splits into five boxes of the
following dimensions: $(M_{J}=1/2)\rightarrow \Gamma _{7}(a)\rightarrow
\left( 7\times 7\right) $, $(M_{J}=3/2)\rightarrow \Gamma _{9}(a)\rightarrow
\left( 6\times 6\right) $, $(M_{J}=5/2)\rightarrow \Gamma _{8}(a)\rightarrow
\left( 4\times 4\right) $, $(M_{J}=7/2)\rightarrow \Gamma _{8}(b)\rightarrow
\left( 2\times 2\right) $, $(M_{J}=9/2)\rightarrow \Gamma _{9}(b)\rightarrow
\left( 1\times 1\right) $, and there are no matrix elements of $M_{J}=11/2$,
i.e. $\Gamma _{7}(b)$, within the subspace $\{^{4}F,^{4}P\}$ of $d^{7}$ that
corresponds to $S=3/2$. The matrices we have obtained coincide with those
given in table II of reference \onlinecite{Palacios}, and their eigenvalues
have been calculated for the different sets of $B$, $D_{s}$ and $D_{t}$
values that were obtained above, employing the one-electron spin-orbit
parameter $\zeta =580\ {\rm cm}^{-1}$. For all the set of parameters in
table \ref{T6p} the lowest doublet is a $\Gamma _{7}(a)$ ($M_{J}=\pm 1/2$),
separated by at least $75\ {\rm cm}^{-1}$ from the following $\Gamma _{9}(a)$
($M_{J}=3/2$) doublet, and by more than $2377\ {\rm cm}^{-1}$ from the
remaining doublets. This situation is not altered by making fairly large
changes in the three basic parameters $B$, $D_{s}$ and $D_{t}$, and shows
that even for moderate increases in the temperature only the lowest doublet (%
$M_{J}=\pm 1/2$) would be occupied. This doublet has allowed ESR
transitions, and should be observed within the approximation employed. If
the position of the two lowest doublets were exchanged, the ESR transitions
of the lowest doublet would be forbidden and the spectra should not be then
observed.

The fact that the two lowest doublets have $M_{J}=\pm 1/2$ and $M_{J}=\pm
3/2 $ and are separated by a large energy from the remaining doublets is
easily understood when we notice that the lowest level in the absence of
spin-orbit interaction is $^{4}A_{2}^{\prime }$. The orbital part $%
A_{2}^{\prime }$ is a singlet with no orbital angular momentum, and the
total ${\bf J}$ would then correspond to the $S=3/2$. These four states
would be rather far apart from the remaining ones, and would split in the
way calculated above through the higher order spin orbit mixing with those
excited states.

The present calculation was for a perfect trigonal bipyramid with $D_{3h}$
symmetry, and one wonders whether the deformations with respect to this
structure could alter the relative position of the two lowest doublets, thus
changing from an allowed to a forbidden ESR transition. We shall then study
the effect of these deformations, both on the relative position of the two
lowest doublets and on the value of the gyromagnetic tensor. In this study
we shall follow a treatment similar to that employed in the octahedral case,
by considering the effect of the normal modes of the trigonal bipyramid on
the Hamiltonian of the penta-coordinated {\rm Co}$^{2+}$.

\subsubsection{The normal modes of the trigonal bipyramid\label{lb}}

\label{B3}As in the octahedral case we are interested in a contribution to
the Hamiltonian of the same type of Eq. (\ref{Et3}), but here the normal
modes $Q_{j}$ and$\ V_{j}({\bf r})$ transform like the same partners of
irreducible representations of the trigonal bipyramid. As the undistorted
complex does not have a center of symmetry, both the even an odd modes
against reflection in the equatorial plane may have non-zero matrix elements
inside the configuration $d^{7}$ of {\rm Co}$^{2+}$, and therefore we shall
need to consider both types of normal modes in our discussion.

The departures of the six atoms of the complex span a reducible
representation $\Gamma $ of the $D_{3h}$ group, that can be reduced as
follows: $\Gamma =2A_{1}^{\prime }+A_{2}^{\prime }+4E^{\prime
}+3A_{2}^{\prime \prime }+2E^{\prime \prime }$ (see e.g. the Eq. (9.19) in
reference \onlinecite{Woodward}). Of these irreducible representations, the $%
A_{2}^{\prime }$ corresponds to an axial rotation, one $E^{\prime \prime }$
to two equatorial rotations, one $A_{2}^{\prime \prime } $ to an axial
translation and one $E^{\prime }$ to two equatorial translations. After
eliminating these three translations and rotations we are left with three
even irreducible representations $E^{\prime }$, as well as two $%
A_{2}^{\prime \prime }$ and one $E^{\prime \prime }$ odd representations.
The six even normal modes ($Q_{1},...,Q_{6}$) transform in pairs like the
partners of $E^{\prime }$; they have been obtained employing standard
techniques\cite{Woodward,WilsonDC} and are defined in table \ref{T7}. In the
same way the two modes $A_{2}^{\prime \prime }$ ($Q_{7},Q_{8}$) and the two
partners of $E^{\prime \prime }$ (($Q_{9},Q_{10}$) have been obtained, and
are defined in the same table.

The modes $Q_{3},Q_{4}$, and $Q_{8}$ are translations of only the two axial
oxigens, and the $Q_{5},Q_{6}$, and $Q_{7}$ are translations of only the
three equatorial oxigens while $Q_{9}$ and $Q_{10}$ are rotations along the $%
x$ and $y$ axis of the two axial oxigens. As these modes are only partial
rotations or translations they are capable of changing the crystal field.
The $x$ and $y$ rotations of the three equatorial oxigens can be combined
with $Q_{9}$ and $Q_{10}$ to give full rotations of the trigonal bipyramid,
and the $z$ rotation of the three equatorial oxigens is already a full
rotation, so these three sets of displacements would not appear in our
calculation.

>From tables \ref{T4p}, \ref{T5}, and \ref{T4} we can calculate the
displacements {\bf u}$_{j}$ of the five {\rm O} with respect to their
positions in the reference trigonal bipyramid (cf. section \ref{B1}) and
calculate the corresponding normal modes $Q_{j}$ defined in table \ref{T7}.
Employing these $Q_{j}$ we calculate in the next two sections the g values
of the distorted complex.

\subsubsection{The effect of the normal modes on the crystal field\label{B4}}

We can now try and find an expression similar to Eq.~(\ref{Et3}) for the
trigonal bipyramid. To this purpose we have employed a relation equivalent
to Eq. (\ref{E2.2}) to calculate, for each of the ten normal modes $Q_{j}$
given in table \ref{T7}, the change in the crystal field $V_{cf}({\bf r})$
when all the ligands are displaced from their equilibrium position in the
reference complex by a small fraction $\varepsilon $ of that particular
normal mode $Q_{j}$. Expanding this change of $V_{cf}({\bf r})$ in a power
series of the coefficient $\varepsilon $ and taking the linear terms in $%
\varepsilon $ gives the corresponding $V_{j}({\bf r})$ from Eq.~(\ref{Et3}).
As we are only interested in the subspace $\{^{4}F,^{4}P\}$ with $S=3/2$ of
the configuration $d^{7}$, and the $V_{j}({\bf r})$ are independent of the
spin component $M_{S}$, we need a $10$x$10$ matrix $\langle
^{4}L,M_{S},M_{L}|V_{CF}^{\prime }|^{4}L^{\prime },M_{S},M_{L}^{\prime
}\rangle $ for each $Q_{j}$, with fixed $M_{S}$ and $L,L^{\prime }=3,1$.
There are regularities between the matrix elements associated to different $%
Q_{j}$, and we shall employ the following abbreviations: $%
Q_{a}=(Q_{2}+iQ_{1})+5(Q_{6}+iQ_{5})$, $%
Q_{b}=3(Q_{2}+iQ_{1})+7(Q_{6}+iQ_{5}) $, $%
Q_{c}=9(Q_{2}+iQ_{1})+(Q_{6}+iQ_{5})$, $Q_{d}=iQ_{7}$, and $%
Q_{e}=9(Q_{10}+iQ_{9})$, as well as their complex conjugates $Q_{a}^{\ast }$%
, $Q_{b}^{\ast }$ $Q_{c}^{\ast }$, $Q_{d}^{\ast }$ and $Q_{e}^{\ast }$. In
tables \ref{T8} and \ref{T10} we give the non-zero matrix elements in the
upper triangle of the submatrices $^{4}F$~x~$^{4}F$ and $^{4}P$~x~$^{4}P$
respectively, and in table \ref{T9} we give all those associated with $^{4}P$%
~x~$^{4}F$; the remaining non-zero elements are obtained by Hermitian
conjugation. It is interesting to note that the matrix elements associated
to $Q_{3}$, $Q_{4}$, and $Q_{8}$ are all zero: these modes involve only the
two axial ions, and the corresponding two atom partial complex is not only
invariant against the operations of $D_{3h}$, but also against a twofold
axis along the $z$ direction. This extra symmetry forces all the
one-electron matrix elements between $d$ states of the crystal field
associated to $Q_{3}$, $Q_{4}$, and $Q_{8}$ to be zero.

As in the crystal field of the reference complex, the $V_{CF}^{\prime }$ has
coefficients containing the {\rm Co-O} distances $R_{a}$ and $R_{c}$, as
well as the atomic averages $\langle r^{2}\rangle $ and $\langle
r^{4}\rangle $, and they appear as $c_{2}$ and $c_{4}$ in the tables \ref{T8}%
, \ref{T9} and \ref{T10}. As the $R_{a}$ and $R_{c}$ are nearly the same, it
is possible from Eq. (\ref{E2.1}) to relate the crystal field parameters $%
D_{s}$ and $D_{t}$ to these two coefficients. Assuming that $R_{a}=R_{c}$ we
obtain $c_{2}=14\ D_{s}$ and $c_{4}=(168/25)D_{t}$, but we have derived
slightly better relations considering the difference between $R_{a}$ and $%
R_{c}$:
\begin{eqnarray}
c_{2} &=&\frac{7}{-(3/2)+2\ (R_{a}/R_{c})^{3}}D_{s},  \nonumber \\
c_{4} &=&\frac{21}{(9/8)+2\ (R_{a}/R_{c})^{5}}D_{t}.  \label{E2.3}
\end{eqnarray}

As with the reference complex, we employ the 3-j coefficients to calculate
the matrix elements of the crystal field $V_{CF}^{\prime }$ in the
representation that diagonalizes the total $J$ and $J_{z}$, because the
doublets $\left| d^{7}\alpha SLJM_{J} \right\rangle $ are basis for the
irreducible representations of the reference trigonal bipyramid, and the
eigenstates of the reference complex would then belong to subspaces with
fixed $M_{J} $. In section \ref{B2} we have shown that, with the two sets of
parameters $B,$ $D_{s}$ and $D_{t}$ obtained in that section, the two lowest
doublets belong to the $M_{J} =1/2$ and $M_{J} =3/2$ subspaces and that they
are separated by more than $75\ {\rm cm}^{-1}$, while the remaining doublets
are more than $2300\ {\rm cm}^{-1}$ above them. A good approximation to
calculate the effect of $V_{CF}^{\prime }$ on these levels is then to
consider the total Hamiltonian inside the two subspaces $M_{J} =1/2,3/2$,
and one has then to consider a matrix of 26 x 26 elements, corresponding to
values of $J$ equal to $9/2,...,1/2$. The eigenstates of this matrix show
that there is no change in the relative position of the two lowest doublets,
so that the ground state remains $M_{J} =1/2$.

\subsubsection{The g-factors of the penta-coordinated {\rm Co}$^{2+}$. \label%
{B5}}

To calculate the spin Hamiltonian we employ the traditional
method\cite{Pryce}. In the present case we consider the four
states of the two lowest doublets of the reference trigonal
bipyramid calculated in section \ref{B2} as the eigenstates of the
unperturbed Hamiltonian, with $M_{J} =1/2$ as the ground doublet
and $M_{J} =3/2$ as the excited one. Both the Zeeman term and the
crystal field $V_{CF}^{\prime }$ produced by the deformation of
the normal modes are the perturbations, and in the usual way we
find the gyromagnetic tensor ${\bf g}$ in second order. We have
calculated the three components of ${\bf g}$ for the
penta-coordinated {\rm Co}$^{2+}$ for all the sets of $B$, $D_{s}$
and $D_{t}$ given in table \ref{T6p}, and the results are given in
table \ref{T11}. The values corresponding to the reference
trigonal bipyramid are given in the rows $b_{0}$ and $c_{0}$,
while those given in the rows $b_{j}$ and $c_{j}$ (with $j=1,2$
for the phosphate and $j=1,2,3$ for the arsenate) have been
calculated employing the normal modes $Q_{j}$ derived from the
crystallographic positions (cf. table \ref{T4p}) as discussed in
section \ref{B3}. One verifies in table \ref{T11} that the average
of the principal values of ${\bf g}$ is not very different from
4.33, but that it changes with $B$ and the crystal field
parameters more than in the octahedral case. The rows $b_{1}$ and
$c_{1}$ include the effect of all normal modes, while in $b_{2}$
and $c_{2}$ only the even modes are considered. In the phosphate
case, the crystallographic $Q_{1}=Q_{5}=0$, and for the arsenate
we have also imposed this condition in rows $b_{3}$ and $c_{3}$.
>From table \ref{T11} we conclude that

\begin{itemize}
\item  With only even modes and $Q_{1}=Q_{5}=0$ (compare rows $b_{2}$ and $%
c_{2}$ in phosphate and $b_{3}$ and $c_{3}$ in arsenate with rows $b_{1}$
and $c_{1}$) the average $g$ is not altered by the lower symmetry crystal
fields generated by the remaining normal modes, and these fields affect the
equatorial components of ${\bf g}$, but leave their sum and the axial
component unaffected.

\item  Only the axial $g$-factor is altered by the inclusion of the odd
modes, while the two equatorial $g$-factors are not altered (compare rows $%
b_{1}$ with $b_{2}$ and $c_{1}$ with $c_{2}$). This result is true both with
$Q_{1}=Q_{5}=0$ (in the phosphate) or otherwise (in the arsenate).

\item  The fields associated to the modes $Q_{1}$ and $Q_{5}$ change the sum
of the two equatorial $g$-factors but leave the axial value unaffected
(compare rows $b_{2}$ with $b_{3}$ and $c_{2}$ with $c_{3}$ in the arsenate).
\end{itemize}

>From the calculations in the present section, it follows that one should
observe an allowed ESR line of {\rm Co}$^{2+}$ from the penta-coordinated
complex when that site is occupied. We have seen in Section \ref{secIIC}
that besides the lines associated to the octahedral spectra, there are some
weak extra lines that could be interpreted as belonging to that complex:
their estimated $g$-factors are given in rows b) of table~\ref{T01}, and
should be compared with the values given in table \ref{T11}, that were
calculated for different sets of parameters derived from the optical spectra
and with normal modes calculated from the crystallographic positions. It is
clear that the arsenate values in row $c_{3}$ of table \ref{T11} are fairly
close to the estimated values in row $b$ from table~\ref{T01}. It is well
known that the $g(i)$ obtained from the crystallographically calculated
normal modes are generally different from those experimentally observed, as
discussed for the octahedral compounds (cf. section \ref{ssec:A} ), and we
could expect that a good fitting could be obtained by making small changes
in the crystallographic normal modes. To verify this assumption it is
sufficient to change only $Q_{2}$ and $Q_{6}$, keeping all the remaining
modes at their crystallographic values. Employing $Q_{2}/R_{a}=-0.03$ and $%
Q_{6}/R_{a}=-0.07$ for the phosphate we find $g(1)=7.05$, $g(2)=3.03$, and $%
g(3)=2.12$, while for the arsenate we obtain $g(1)=7.62$, $g(2)=3.04$, and $%
g(3)=1.99$ with $Q_{2}/R_{a}=0.01$ and $Q_{6}/R_{a}=-0.07$. These fittings
are fairly good, and show that the ESR spectra of {\rm Co}$^{2+}$ in the two
compounds can be described perfectly well within our theory, but the crystal
fields obtained can not be taken too seriously because of the very large
errors in the experimental ${\bf g}$ tensor.

We notice that the relative intensities of the extra lines in Fig.~\ref
{Fig01} are rather smaller than those in Fig.~\ref{Fig02}. This can be
understood because the concentration of {\rm Co}$^{2+}$ in the arsenate is
ten times larger than in the phosphate, and this should also alter their
relative occupations.

The rather small intensity of the lines that could be attributed to the
penta-coordinated complex indicates a very small occupation of {\rm Co}$%
^{2+} $ in the penta-coordinated sites. To verify this conjecture, we
present a molecular orbital calculation of the heat of formation of these
compounds in the following section, and the results are compatible with the
present conclusion.

\subsection{Molecular Orbital Calculations}

The energetics of penta- and hexa-coordinated phosphate clusters
has been assessed in terms of the molecular orbital theory. Heats
of formation were calculated within the well known semi-empirical
technique Parametric Model 3 (PM3)\cite{pm3}. This is a technique
derived from the Hartree-Fock approximation in combination with a
minimal basis set expansion of the molecular orbitals. Here we
used a special parametrization developed for transition metal
atoms which is contained in the package SPARTAN\cite{spartan}.
Correlation and relativistic effects, which are not explicitly
treated in this theory, are partly recovered from the adoption of
experimental data in the parametrization. Molecular geometries
were obtained as follows: the central metal atom and the
coordinates of the first neighboring five or six oxygen atoms were
taken from the crystal structure of the compound {\rm
Co}$_{2}${\rm (OH)PO}$_{4}$. The ligands were chosen to be
phosphoric acid molecules, {\rm OP(OH)}$_{3}$ , since they have
all bonds saturated and are neutral. Geometry optimizations of the
isolated ligand were carried out at the {\it ab-initio } 6-31G**
level of calculation, in which the atomic orbitals of the basis
set are written as a linear combination of cartesian gaussian
functions\cite{basis}. The proton-free oxygen atoms of the ligands
were then placed in the crystallographic positions of the oxygens
around the metal atom such that the O=P bond points to the M-O
direction, as shown in Figs.~\ref{Fig1} and \ref{Fig2}. The PM3
heats of formation of the clusters {\rm [M(OP(OH)}$_{3}${\rm )}$_{n}${\rm ]}$%
^{2+}$, {\it n} = 5,6 and {\rm M} = Co and Zn\ were calculated assuming that
Zn ion just replace the Co ion at frozen ligands positions. This is a
reasonable assumption since the pure Co and Zn crystals have very similar
cell parameters. Co clusters are doublets so that the unrestricted
(spin-polarized) PM3 Hamiltonian was adopted. Spin contamination was
negligible in this calculation. In order to discount the energies associated
to the ligands themselves, the heats of formation of the corresponding
clusters without the central metal ion were computed. Results are shown in
table~\ref{T13}. The values in the first column are the contributions from
the ligands to the metal clusters heats of formation. It is then possible to
evaluate the relative stabilization of Co$^{2+}$ and Zn$^{2+}$ ions in the
penta- and hexa-coordinated environments by making the difference between
the values in columns two or three and column one. It gives the energies $%
-1258.04$ kcal/mol and $-1314.30$ kcal/mol, for Co$^{2+}$ at the trigonal
bipyramid and octahedral sites respectively, while for Zn$^{2+}$ the values
are 346.19 kcal/mol and 340.00 kcal/mol. These values show that Co prefers
the octahedral site by an amount of \ $\sim $56 kcal, which is approximately
2.4 eV, and Zn is slightly more stable also at the octahedral site by \ $%
\sim $6 kcal, or 0.3 eV. This difference is due to the partially filled 3d
orbitals of Co that interact with the lone pairs of the neighboring oxygens,
giving a more covalent character to the interaction.

A more direct comparison was made through the calculation of the heat of
formation of clusters where the metal atoms and first neighbors are in the
conformation of the {\rm Co}$_{2}${\rm (OH)PO}$_{4}$\ unit cell, as
illustrated in Fig~\ref{Fig3}. Two clusters were built such that in one the
Co$^{2+}$ ion occupies the hexa-coordinated site and Zn$^{2+}$ ion is in the
penta-coordinated site (cluster 1), while in the other (cluster 2) Co$^{2+}$
and Zn$^{2+}$ ions are exchanged. The phosphate ions in contact with both
metal ions were substituted by {\rm H}$_{2}${\rm PO}$_{4}$\ species and the
remainder of the ligands were phosphoric acid molecules The composition of
these clusters is then {\rm [CoZn(OH)(H}$_{2}${\rm PO}$_{4}${\rm )}$_{2}$%
{\rm (OP(OH)}$_{3}${\rm )}$_{5}${\rm ]}$^{4+}$. PM3 calculations gave $%
\Delta H_{f}$ (cluster 1) = $-2147.10$ kcal/mol and $\Delta H_{f}$ (cluster
2) = $-2094.44$ kcal/mol, that is, the cluster with Co$^{2+}$ ion in the
octahedral site is \ $\sim $53 kcal more stable than the other one.\ It is
thus expected that Co impurities in the zinc compounds occupy preferentially
the octahedral sites, and this conclusion agrees with very low intensity of
the ESR lines attributed to Co$^{2+}$ in the penta-coordinated sites of the
dilute compounds, as discussed in the previous section.

\section{DISCUSSION AND CONCLUSIONS}

\label{secIV}

Four compounds of the adamite type: {\rm Zn}$_{2}${\rm (OH)PO}$_{4}$, {\rm Mg%
}$_{2}{\rm (OH)AsO}_{4}$, {\rm Co}$_{2}${\rm (OH)PO}$_{4}$, and {\rm Co}$_{2}%
{\rm (OH)AsO}_{4}$.have been synthesized and studied, and the measurement of
the optical properties of the pure {\rm Co }compounds and of the ESR of
impurities of {\rm Co}$^{2+}$ in {\rm Zn}$_{2}${\rm (OH)PO}$_{4}$ and {\rm Mg%
}$_{2}{\rm (OH)AsO}_{4}$ have been discussed. Crystal field theory has been
employed to try and understand the experimental ESR results for the two {\rm %
Co}$^{2+}$ complexes with coordination five and six that are present in the
adamite structure. The Racah parameter $B$ as well as the crystal fields $%
D_{q}$ for the octahedral complex and both $D_{s}$ and $D_{t} $ for the
trigonal bipyramid one have been estimated from the assignments that were
made of the diffuse reflectance spectrum of these two complexes. Two
alternative sets of parameters were proposed for the penta-coordinated
complex.

>From the crystallographic structure, a reference octahedron centered in the\
{\rm Co}$^{2+}$ was defined, such that the normal modes of the complex
corresponding to rotations and expansions would be zero and the remaining
normal modes would not have any contribution of these irrelevant
deformations. Using a method already applied\cite{GoniFB} to study the ESR
of {\rm Co}$^{2+}$ in {\rm NH}$_{4}${\rm NiPO}$_{4}${\rm .6H}$_{2}${\rm O},
the crystal fields that would reproduce the experimental ${\bf g}$ tensor of
the octahedral complex in both {\rm Zn}$_{2}${\rm (OH)PO}$_{4}$ and {\rm Mg}$%
_{2}{\rm (OH)AsO}_{4}$ have been obtained.

As the penta-coordinated complex seems to have at most minor contributions
to the ESR spectra, we have analyzed the possible motives for this behavior.
We argue that two doublets with $M_{J} =\pm 1/2$ and $M_{J} =\pm 3/2$ would
be lowest in energy, separated by a rather large excitation energy from the
remaining excited states. The $M_{J} =\pm 3/2$ has forbidden ESR
transitions, and this would explain the experimental results if that were
the ground doublet, but when the crystal field of the trigonal bipyramid is
considered together with the spin-orbit interaction, it was found that the $%
M_{J} =\pm 1/2$ doublet, with allowed ESR transitions, is the lowest. To
verify whether this result would be altered by the deformations of the
trigonal bipyramid, we considered their effects in a way similar to that
employed in the octahedral case to calculate the ${\bf g}$ tensor. First it
was necessary to derive the normal modes of the trigonal bipyramid that are
relevant to our problem, and they are given in table \ref{T7}. The
corresponding Jahn-Teller contributions $V_{CF}^{\prime }$ to the crystal
field, whose non-zero matrix elements $\langle
^{4}L,M_{S},M_{L}|V_{CF}^{\prime }|^{4}L^{\prime },M_{S},M_{L}^{\prime
}\rangle $,for $L,L^{\prime }=3,1$, are given in tables \ref{T8}, \ref{T9}
and \ref{T10}. Defining a reference perfect trigonal bipyramid by the same
method employed in the octahedral case, the values of the relevant normal
modes were obtained by employing the crystallographic positions, and
subsequently used to calculate their effect on the relative position of the
two lowest doublets. No appreciable change was found, and as an alternative
explanation we assumed that the penta-coordinated complex is scarcely
occupied in the dilute system. To verify this conclusion, the heat of
formation of the octahedral and the trigonal bipyramid complexes with both
{\rm Co} and {\rm Zn} as the central ions were calculated, and it was found
that their values are compatible with a rather small occupation of the
penta-coordinated site.

Employing the Jahn-Teller crystal fields together with the normal modes
calculated from the crystallographic distortions, it was possible to
calculate the ${\bf g}$ tensor, shown in table \ref{T11} for both the
perfect and deformed trigonal bipyramid, this last subjected to different
deformations. The trace of the ${\bf g}$ tensor in the perfect trigonal
bipyramid changes more with the parameters $B$,$D_{s} $ and $D_{t}$ than in
the octahedral case, where it is always fairly close to 13.

The trigonal bipyramid has no center of symmetry, and it was necessary to
consider all the normal modes, even those that are odd against reflection in
the horizontal symmetry plane. We have shown that these last modes affect
the axial component of ${\bf g}$ but that they have little or no effect on
the two equatorial components. The experimental $g$-tensor of the
penta-coordinated complex could be measured but with rather large errors. As
in the octahedral case, the crystallographically determined normal modes
could not explain the observed values, but for the two type of complexes it
was possible to find values of the normal modes that would generate crystal
fields that describe the experimental ESR spectra for both the phosphate and
arsenate compounds.

We conclude that both our theoretical analysis of the ESR of the systems
studied, as well as the molecular orbital calculation of the formation
energies. coincide in assigning a rather small relative occupation of the
penta-coordinated sites with respect to the octahedral ones in those
systems. We could also understand the experimental spectra of both the
octahedral and penta-coordinated complexes by considering the effect of the
crystal fields generated by the corresponding normal modes on the ${\bf g}$
tensor.

\acknowledgements

The three first authors(MEF,MCS,GEB) would like to acknowledge financial
support from the following agencies: FAPESP and CNPq .

\begin{table}[tbp]
\centering
\par
\begin{tabular}{rrrrrr}
\hline
$^{4}${\bf T}$_{1g}\rightarrow $ & $^{4}${\bf T}$_{2g}$ & $^{4}${\bf A}$%
_{2g} $ & $^{4}${\bf T}$_{1g}$(P) & ${\bf B}$ & ${\bf D}_{q}$ \\ \hline\hline
&  & ${\rm PO}_{4}$ &  &  &  \\ \hline
a) & 8450 & 15450 & 18350 &  &  \\
b) & 7819 & 16013 & 18324 & 767.6 & 819.4 \\ \hline
&  & ${\rm AsO}_{4}$ &  &  &  \\ \hline
a) & 7700 & 15500 & 18020 &  &  \\
b) & 7616 & 15585 & 18011 & 758.9 & 796.9 \\ \hline
\end{tabular}
\caption{a) The transitions between the ground $^{4}T_{1g}$ level and the
levels shown at the top of each column, in {\rm cm}$^{-1}$ and assigned from
the experimental spectra of the octahedral complexes of {\rm Co}$_{2}{\rm %
(OH)PO}_{4}$ and {\rm Co}$_{2}{\rm (OH)AsO}_{4}$; the level $^{4}T_{1g}(P)$
corresponds to the highest of the same symmetry. b) The best fit, obtained
with the $B$ and $D_{q}$ shown in the last two columns. }
\label{T1p}
\end{table}

\begin{table}[tbp]
\centering
\par
\begin{tabular}{cccccc}
\hline
$^{4}A_{2}^{\prime }\rightarrow $ & $^{4}A_{1}^{\prime \prime }$, $%
^{4}A_{2}^{\prime \prime }$ & $^{4}E^{\prime \prime }$ & $^{4}E^{\prime }$ &
$^{4}A_{2}^{\prime }(P)$ & $^{4}E^{\prime \prime }(P)$ \\ \hline\hline
&  &  & ${\rm PO}_{4}$ &  &  \\ \hline
a) & 6400 & 7000 & 11100 & 15800 & 19600 \\
b) & 3233 & 4835 & 12868 & 17386 & 17947 \\
c) & 1511 & 3603 & 11106 & 15801 & 19604 \\ \hline\hline
&  &  & ${\rm AsO}_{4}$ &  &  \\ \hline
a) & 5000 & 6250 & 10870 & 16000 & 19800 \\
b) & 2707 & 4440 & 12210 & 17188 & 18595 \\
c) & 1437 & 3535 & 10876 & 15999 & 19805 \\ \hline
\end{tabular}
\caption{a) The transitions between the ground level $^{4}A_{1}^{\prime%
\prime }$ and the levels shown at the top of each column, given in {\rm cm}$%
^{-1}$ and assigned from the experimental spectra of the penta-coordinated
complexes of {\rm Co}$_{2}{\rm (OH)PO}_{4}$ and {\rm Co}$_{2}{\rm (OH)AsO}%
_{4}$. b) The best possible fit to the five transitions. c) The best fit
obtained by adjusting only the three transitions of higher energy. The
corresponding values of $B$, $D_{s}$, $D_{t}$ are given in rows b) and c) of
table \ref{T6p}. }
\label{T6}
\end{table}

\begin{table}[tbp]
\centering
\par
\begin{tabular}{ccccccc}
\hline
& g$_{1}$ & g$_{2}$ & g$_{3}$ & A$_{1}$ & A$_{2}$ & A$_{3}$ \\ \hline\hline
&  &  & PO$_{4}$ &  &  &  \\
a & 5.89$\pm 0.02$ & 4.55$\pm 0.05$ & 2.02$\pm 0.02$ & 240$\pm 5$ & 155$\pm
8 $ & 85$\pm 3$ \\
b & 8. $\pm 0.5$ & 3.2 $\pm 0.3$ & 2.0 $\pm 0.2$ &  &  &  \\ \hline
&  &  & AsO$_{4}$ &  &  &  \\
a & 6.22$\pm 0.02$ & 4.21$\pm 0.05$ & 2.05$\pm 0.02$ & 140$\pm 5$ & 120$\pm
7 $ & 55$\pm 5$ \\
b & 9. $\pm 1.5$ & 3. $\pm 0.5$ & 2.0 $\pm 0.2$ &  &  &
\end{tabular}
\caption{Values of the principal g and A parameters, obtained from the
spectra in Figs.~\ref{Fig01} and \ref{Fig02}. The values of the A parameters
are in $10^{-4}$ cm$^{-1}$ units. a) Octahedral complex: the g and A values
were obtained from a program simulating powder spectra, as described in the
text. b) The parameters for the penta-coordinated {\rm Co}$^{2+}$, also
estimated by simulation.}
\label{T01}
\end{table}

\begin{table}[tbp]
\centering
\par
\begin{tabular}{rrrrrrr}
\hline
{\bf n} & {\bf a} (\AA ) & {\bf b} (\AA ) & {\bf c} (\AA ) & {\bf a} (\AA )
& {\bf b} (\AA ) & {\bf c} (\AA ) \\ \hline\hline
&  & ${\rm PO}_{4}$ &  &  & ${\rm AsO}_{4}$ &  \\ \hline
1 & 4.7150 & $-1.2420$ & 0.0000 & \ 0.9197 & $-1.0794$ & 0.0000 \\
2 & 4.9627 & $-1.0327$ & 2.9700 & \ 0.6438 & $-1.2633$ & 3.0255 \\
3 & 2.1592 & $-1.2713$ & 1.7274 & $-1.9091$ & $-1.1662$ & 1.3476 \\
4 & 3.0793 & \ 1.0327 & 2.9700 & $-0.6438$ & \ 1.2633 & 3.0255 \\
5 & 3.3270 & \ 1.2420 & 0.0000 & $-0.9197$ & \ 1.0794 & 0.0000 \\
6 & 5.8827 & \ 1.2713 & 1.7274 & \ 1.9091 & \ 1.1662 & 1.3476 \\
7 & 4.0210 & \ 0.0000 & 1.5135 & \ 0.0000 & \ 0.0000 & 1.4910 \\ \hline
\end{tabular}
\caption{The columns {\bf a,b,c} give the position of the six oxygens
(n=1,...,6) and Cobalt (n=7) in the hexa-coordinated complexes of {\rm Co}$%
_{2}{\rm (OH)PO}_{4}$ and {\rm Co}$_{2}{\rm (OH)AsO}_{4}$ with respect to
the three unit cell axes.The X,Y,Z axes roughly correspond to n=1,2,3
respectively, taking n=7 as the origin.}
\label{T1}
\end{table}

\begin{table}[tbp]
\centering
\par
\begin{tabular}{rrrrrr}
\hline
& {\bf Q}$_{2}/R$ & {\bf Q}$_{3}/R$ & {\bf Q}$_{4}/R$ & {\bf Q}$_{5}/R$ &
{\bf Q}$_{6}/R$ \\ \hline\hline
&  &  & ${\rm PO}_{4}$ &  &  \\ \hline
a) & 0 & $-0.04407$ & 0.03478 & 0.03478 & 0.11425 \\
b) & 0 & $-0.11683$ & $-0.01717$ & $-0.01717$ & $-0.05641$ \\ \hline
&  &  & ${\rm AsO}_{4}$ &  &  \\ \hline
a) & 0 & $-0.0441$ & 0.04032 & 0.04032 & 0.09989 \\
b) & 0 & $-0.09328$ & $-0.02818$ & $-0.02818$ & $-0.06982$ \\ \hline
\end{tabular}
\caption{Normal modes of the octahedral Co divided by the Co-O distance R in
{\rm Co:Zn}$_{2}{\rm (OH)PO}_{4}$ and {\rm Co:Mg}$_{2}{\rm (OH)AsO}_{4}$. a)
Values that adjust the experimental values of the g tensor. b) Values
obtained from the crystallographic positions corresponding to the pure
compounds. }
\label{T2}
\end{table}

\begin{table}[tbp]
\centering
\par
\begin{tabular}{rrrrrrr}
\hline
& {\bf X} & {\bf Y} & {\bf Z} & {\bf X} & {\bf Y} & {\bf Z} \\ \hline\hline
&  & ${\rm PO}_{4}$ &  &  & ${\rm AsO}_{4}$ &  \\ \hline
{\bf a} & 0.4054 & $-0.5794$ & $-0.7071$ & 0.3800 & $-0.5963$ & $-0.7071$ \\
{\bf b} & 0.4054 & $-0.5794$ & $\ \ 0.7071$ & 0.3800 & $-0.5963$ & $\ \
0.7071$ \\
{\bf c} & $-0.8194$ & $-0.5733$ & 0.0000 & $-0.8434$ & $-0.5373$ & 0.0000 \\
\hline
\end{tabular}
\caption{Direction cosines of the three axis {\bf X,Y,Z} of the reference
perfect octahedron of {\rm Co}$_{2}{\rm (OH)PO}_{4}$ and {\rm Co}$_{2} {\rm %
(OH)AsO}_{4}$ with respect to the three crystallographic axis {\bf a, b, c}}
\label{T3}
\end{table}

\begin{table}[tbp]
\centering
\par
\begin{tabular}{rrrrrrr}
\hline
{\bf n} & {\bf a} (\AA ) & {\bf b} (\AA ) & {\bf c} (\AA ) & {\bf a} (\AA )
& {\bf b} (\AA ) & {\bf c} (\AA ) \\ \hline\hline
&  & ${\rm PO}_{4}$ &  &  & ${\rm AsO}_{4}$ &  \\ \hline
1 & 4.9008 & 3.2438 & 2.9700 & 4.9744 & 0.9132 & 3.0180 \\
2 & 1.8617 & 2.9132 & 1.2426 & 1.9003 & 1.1604 & 1.3442 \\
3 & 1.8617 & 2.9132 & 4.6974 & 1.9003 & 1.1604 & 4.6676 \\
4 & 3.1412 & 5.1252 & 2.9700 & 3.2085 & 3.2015 & 3.0180 \\
5 & 3.0793 & 1.0327 & 2.9700 & 3.2736 & -0.9132 & 3.0180 \\
6 & 2.9102 & 3.0621 & 2.9700 & 3.0056 & 1.1561 & 3.0180 \\ \hline
\end{tabular}
\caption{The columns {\bf a,b,c} give the position of the five oxygens
(n=1,...,5) and Cobalt (n=6) in the penta-coordinated complexes of {\rm Co}$%
_{2}{\rm (OH)PO}_{4}$ and {\rm Co}$_{2}{\rm (OH)AsO}_{4}$ with respect to
the three unit cell axes.}
\label{T4p}
\end{table}

\begin{table}[tbp]
\centering
\par
\begin{tabular}{lllllll}
\hline
& {\bf X} & {\bf Y} & {\bf Z} & {\bf X} & {\bf Y} & {\bf Z} \\ \hline\hline
&  & ${\rm PO}_{4}$ &  &  & ${\rm AsO}_{4}$ &  \\ \hline
{\bf a} & \ \ 0. & 0. & 1. & -0.00203 & \ \ 0. & 0.999998 \\
{\bf b} & \ \ 0.99931 & 0.03720 & 0. & \ 0.999698 & $-0.02449$ & 0.00203 \\
{\bf c} & $-0.03720$ & 0.99931 & 0. & \ 0.02449 & \ \ 0.9997 & 0.00005 \\
\hline
\end{tabular}
\caption{Direction cosines of the three axis {\bf X,Y,Z} employed to define
the reference perfect trigonal bipyramid of {\rm Co}$_{2}{\rm (OH)PO}_{4}$
and {\rm Co}$_{2} {\rm (OH)AsO}_{4}$ with respect to the three
crystallographic axis {\bf a, b, c}}
\label{T5}
\end{table}

\begin{table}[tbp]
\centering
\par
\begin{tabular}{ccccc}
\hline
{\bf n} & X & Y & Z &  \\ \hline\hline
1 & 0 & R$_{a}$ & 0 &  \\
2 & $-\frac{\sqrt{3}}{2}$$R_{a}$ & $-\frac{1}{2}R_{a}$ & 0 &  \\
3 & $\ \frac{\sqrt{3}}{2}$$R_{a}$ & $\ \frac{1}{2}R_{a}$ & 0 &  \\
4 & 0 & 0 & $\ R_{c}$ &  \\
5 & 0 & 0 & $-R_{c}$ &  \\
6 & 0 & 0 & 0 &  \\ \hline
\end{tabular}
\caption{The columns X,Y,Z give the position of the five oxygens (n=1,...,5)
and Cobalt (n=6) in the reference perfect penta-coordinated complexes of
{\rm Co}$_{2}{\rm (OH)PO}_{4}$ [$R_{a}=2.01622$ \AA \ and $R_{c}=2.04365$
\AA ],and {\rm Co}$_{2} {\rm (OH)AsO}_{4}$ $[R_{a}=1.98578$ \AA \ and $%
R_{c}=2.05596$ \AA ] with respect to the axes defined in table~\ref{T5} }
\label{T4}
\end{table}

\begin{table}[tbp]
\centering
\par
\begin{tabular}{rrrrrrr}
\hline
& {\bf B} & ${\bf D_{s}}$ & ${\bf D_{t}}$ & {\bf B} & ${\bf D_{s}}$ & ${\bf %
D_{t}}$ \\ \hline\hline
&  & ${\rm PO}_{4}$ &  &  & ${\rm AsO}_{4}$ &  \\ \hline
{\bf b} & 728. & 165. & 947. & 785. & 313 & 919. \\
{\bf c} & 852. & 745. & 885. & 875. & 749. & 869. \\ \hline
\end{tabular}
\caption{The values of $B$, $D_{s}$, $D_{t}$ in cm$^{-1}$ that fit the
optical transitions, given in table \ref{T6}, of the two penta-coordinated
complexes. The best fit to the five transitions is given in row b), and the
best fit to the three highest transitions in row c). The spin orbit
parameter $\protect\zeta = 580$ cm$^{-1}$ was used in all these fittings }
\label{T6p}
\end{table}

\begin{table}[tbp]
\centering
\par
\begin{tabular}{cccccccccccccccc}
\hline
Q$_{j}$ & {\bf x}$_{1}$ & {\bf y}$_{1}$ & {\bf z}$_{1}$ & {\bf x}$_{2}$ &
{\bf y}$_{2}$ & {\bf z}$_{2}$ & {\bf x}$_{3}$ & {\bf y}$_{3}$ & {\bf z}$_{3}$
& {\bf x}$_{4}$ & {\bf y}$_{4}$ & {\bf z}$_{4}$ & {\bf x}$_{5}$ & {\bf y}$%
_{5}$ & {\bf z}$_{5}$ \\ \hline\hline
$2\sqrt{3}Q_{1}$ & -$2$ & 0 & 0 & $1$ & $\sqrt{3}$ & $0$ & 1 & -$\sqrt{3}$ &
0 & 0 & 0 & 0 & 0 & 0 & 0 \\
$2\sqrt{3}Q_{2}$ & 0 & 2 & $0$ & $\sqrt{3}$ & -1 & 0 & -$\sqrt{3}$ & -$1$ & 0
& 0 & 0 & 0 & 0 & 0 & 0 \\
$\sqrt{2}Q_{3}$ & 0 & 0 & 0 & 0 & 0 & 0 & 0 & 0 & 0 & 1 & 0 & 0 & 1 & 0 & 0
\\
$\sqrt{2}Q_{4}$ & 0 & 0 & 0 & 0 & 0 & 0 & 0 & 0 & 0 & 0 & 1 & 0 & 0 & 1 & 0
\\
$\sqrt{3}Q_{5}$ & 1 & 0 & 0 & 1 & 0 & 0 & 1 & 0 & 0 & 0 & 0 & 0 & 0 & 0 & 0
\\
$\sqrt{3}Q_{6}$ & 0 & 1 & 0 & 0 & 1 & 0 & 0 & 1 & 0 & 0 & 0 & 0 & 0 & 0 & 0
\\ \hline
$\sqrt{3}Q_{7}$ & 0 & 0 & 1 & 0 & 0 & 1 & 0 & 0 & 1 & 0 & 0 & 0 & 0 & 0 & 0
\\
$\sqrt{2}Q_{8}$ & 0 & 0 & 0 & 0 & 0 & 0 & 0 & 0 & 0 & 0 & 0 & 1 & 0 & 0 & 1
\\
$\sqrt{2}Q_{9}$ & 0 & 0 & 0 & 0 & 0 & 0 & 0 & 0 & 0 & 0 & -1 & 0 & 0 & 1 & 0
\\
$\sqrt{2}Q_{10}$ & 0 & 0 & 0 & 0 & 0 & 0 & 0 & 0 & 0 & 1 & 0 & 0 & -1 & 0 & 0
\\ \hline
\end{tabular}
\caption{The six even (1-6) and four odd (7-10) normal modes $E^{\prime }$
that are relevant to our problem. The numbers are the coefficients of the
departures ${\bf u}_{j}=\{x_{j},y_{j},z_{j}\}$ of the $j$=th ion from their
equilibrium position.}
\label{T7}
\end{table}

\begin{table}[tbp]
\centering
\par
\begin{tabular}{rrc}
\hline
$M_{L}$ & $M_{L}^{\prime }$ & $\langle ^{4}F,M_{S},M_{L}|V_{CF}^{\prime
}|^{4}F,M_{S},M_{L}^{\prime }>\rangle $ \\ \hline\hline
$-3$ & $-2$ & \multicolumn{1}{l}{$-\left( 3\ c2+10\ c4\right) \
Q_{e}^{\ast}/\left( 7\sqrt{3}\right) $} \\
$-3$ & $-1$ & \multicolumn{1}{l}{$\left( 6\ c2\ Q_{a}-15\ c4\
Q_{b}\right)/(56\sqrt{5})$} \\
-$3$ & $0$ & \multicolumn{1}{l}{$\sqrt{15}\ c4\ Q_{d}/4$} \\
-$3$ & $1$ & \multicolumn{1}{l}{-$\sqrt{5}\ c4\ Q_{c}^{\ast }/16$} \\
-$2$ & $-1$ & \multicolumn{1}{l}{-$\left( 9\ c2-40\ c4\right) \
Q_{e}^{\ast}/\left( 21\sqrt{5}\right) $} \\
-$2$ & $0$ & \multicolumn{1}{l}{$\left( 12\ c2\ Q_{a}+5\ c4\ Q_{b}\right)/(56%
\sqrt{10})$} \\
-$2$ & $1$ & \multicolumn{1}{l}{$\sqrt{5}\ c4\ Q_{d}/\left( 2\sqrt{6}\right)$%
} \\
-$2$ & $2$ & \multicolumn{1}{l}{-$5\ c4\ Q_{c}^{\ast }/(16\sqrt{3})$} \\
-$1$ & $0$ & \multicolumn{1}{l}{-$\sqrt{2}\left( 3\ c2-25\ c4\right) \
Q_{e}^{\ast }/\left( 35\sqrt{3}\right) $} \\
-$1$ & $1$ & \multicolumn{1}{l}{$\left( 18\ c2\ Q_{a}+25\ c4\
Q_{b}\right)/(140\sqrt{3})$} \\
-$1$ & $2$ & \multicolumn{1}{l}{-$\sqrt{5}\ c4\ Q_{d}/\left( 2\sqrt{6}%
\right) $} \\
-$1$ & $3$ & \multicolumn{1}{l}{-$\sqrt{5}\ c4\ Q_{c}^{\ast }/16$} \\
$0$ & $1$ & \multicolumn{1}{l}{$\sqrt{2}\left( 3\ c2-25\ c4\right) \
Q_{e}^{\ast }/\left( 35\sqrt{3}\right) $} \\
$0$ & $2$ & \multicolumn{1}{l}{$\left( 12\ c2\ Q_{a}+5\ c4\ Q_{b}\right) /(56%
\sqrt{10})$} \\
$0$ & $3$ & \multicolumn{1}{l}{-$\sqrt{15}\ c4\ Q_{d}/4$} \\
$1$ & $2$ & \multicolumn{1}{l}{$\left( 9\ c2-40\ c4\right) \
Q_{e}^{\ast}/\left( 21\sqrt{5}\right) $} \\
$1$ & $3$ & \multicolumn{1}{l}{$\left( 6\ c2\ Q_{a}-15\ c4\ Q_{b}\right) /(56%
\sqrt{5})$} \\
$2$ & $3$ & \multicolumn{1}{l}{$\left( 3\ c2+10\ c4\right) \
Q_{e}^{\ast}/\left( 7\sqrt{3}\right) $} \\ \hline
\end{tabular}
\caption{ The non-zero matrix elements of the crystal field $V_{CF}^{\prime
}=\sum_{j}Q_{j}\ V_{j}({\bf r})$ generated by the normal modes $Q_{1}$, ...,$%
Q_{10}$ between states $\left| \protect\alpha ,S,M_{S},L,M_{L}\right\rangle
=\left| ^{4}FM_{S}M_{L}\right\rangle $ in the subspace $^{4}F$~x~$^{4}F$.
Only the elements corresponding to the upper triangle of the matrix are
given, and the remaining ones are obtained by Hermitian conjugation. The
matrix is independent of, and diagonal in, the spin components $M_{S}$. To
compress the table we have used the following abbreviations: $%
Q_{a}=(Q_{2}+iQ_{1})+5(Q_{6}+iQ_{5})$, $%
Q_{b}=3(Q_{2}+iQ_{1})+7(Q_{6}+iQ_{5}) $, $%
Q_{c}=9(Q_{2}+iQ_{1})+(Q_{6}+iQ_{5})$, (even modes) and $Q_{d}=iQ_{7}$, $%
Q_{e}=(iQ_{9}+Q_{10})$ (odd modes), as well as their complex conjugates $%
Q_{a}^{\ast }$, $Q_{b}^{\ast }$, $Q_{c}^{\ast }$, $Q_{d}^{\ast }$ and $%
Q_{e}^{\ast }$ }
\label{T8}
\end{table}

\begin{table}[tbp]
\centering
\par
\begin{tabular}{rrr}
\hline
$M_{L}$ & $M_{L}^{\prime }$ & $\langle ^{4}P,M_{S},M_{L}|V_{CF}^{\prime
}|^{4}P,M_{S},M_{L}^{\prime }>\rangle $ \\ \hline\hline
-$1$ & $0$ & \multicolumn{1}{l}{$3\ c2\ Q_{e}^{\ast }/5$} \\
-$1$ & $1$ & \multicolumn{1}{l}{-$3\sqrt{3}\ c2\ Q_{a}/20$} \\
$0$ & $1$ & \multicolumn{1}{l}{-$3\ c2\ Q_{e}^{\ast }/5$} \\ \hline
\end{tabular}
\caption{Same as in table \ref{T8} but for the sub-matrix $^{4}P$ x $^{4}P$.
The same abbreviations are used here.}
\label{T10}
\end{table}

\begin{table}[tbp]
\centering
\par
\begin{tabular}{rrc}
\hline
$M_{L}$ & $M_{L}^{\prime }$ & $\langle ^{4}P,M_{S},M_{L}|V_{CF}^{\prime
}|^{4}F,M_{S},M_{L}^{\prime }\rangle $ \\ \hline\hline
-$1$ & -$3$ & \multicolumn{1}{l}{$\left( 72\ c2\ Q_{a}^{\ast }-5\ c4\
Q_{b}^{\ast }\right) /\left( 56\sqrt{30}\right) $} \\
-$1$ & -$2$ & \multicolumn{1}{l}{$\sqrt{2}\left( 12\ c2\ +\ 5\ c4\right) \
Q_{e}/\left( 7\sqrt{15}\right) $} \\
-$1$ & $0$ & \multicolumn{1}{l}{$-2\left( 18\ c2+25\ c4\right) \ Q_{e}^{\ast
}/105$} \\
-$1$ & $1$ & \multicolumn{1}{l}{$\left( 24\ c2\ Q_{a}-25\ c4\ Q_{b}\right)
/\left( 280\sqrt{2}\right) $} \\
-$1$ & $2$ & \multicolumn{1}{l}{$\sqrt{5}\ c4\ Q_{d}/4$} \\
-$1$ & $3$ & \multicolumn{1}{l}{-$\sqrt{5}\ c4\ Q_{c}^{\ast }/\left( 8\sqrt{6%
}\right) $} \\
$0$ & -$3$ & \multicolumn{1}{l}{$-\sqrt{5}\ c4\ Q_{d}/\left( 4\sqrt{3}%
\right) $} \\
$0$ & -$2$ & \multicolumn{1}{l}{$\left( 12\ c2\ Q_{a}^{\ast }+5\ c4\
Q_{b}^{\ast }\right) /\left( 28\sqrt{10}\right) $} \\
$0$ & -$1$ & \multicolumn{1}{l}{$\sqrt{2}\left( 24\ c2-25\ c4\right) \
Q_{e}/\left( 35\sqrt{3}\right) $} \\
$0$ & $1$ & \multicolumn{1}{l}{-$\sqrt{2}\left( 24\ c2-25\ c4\right) \
Q_{e}/\left( 35\sqrt{3}\right) $} \\
$0$ & $2$ & \multicolumn{1}{l}{$\left( 12\ c2\ Q_{a}+5\ c4\ Q_{b}\right)
/\left( 28\sqrt{10}\right) $} \\
$0$ & $3$ & \multicolumn{1}{l}{$-\sqrt{5}\ c4\ Q_{d}/\left( 4\sqrt{3}\right)
$} \\
$1$ & -$3$ & \multicolumn{1}{l}{-$\sqrt{5}\ c4\ Q_{c}/\left( 8\sqrt{6}%
\right) $} \\
$1$ & -$2$ & \multicolumn{1}{l}{$\sqrt{5}\ c4\ Q_{d}/4$} \\
$1$ & -$1$ & \multicolumn{1}{l}{$\left( 24\ c2\ Q_{a}^{\ast }-25\ c4\
Q_{b}^{\ast }\right) /\left( 280\sqrt{2}\right) $} \\
$1$ & $0$ & \multicolumn{1}{l}{$2\left( 18\ c2+25\ c4\right) \ Q_{e}^{\ast
}/105$} \\
$1$ & $2$ & \multicolumn{1}{l}{-$\sqrt{2}\left( 12\ c2\ +\ 5\ c4\right) \
Q_{e}/\left( 7\sqrt{15}\right) $} \\
$1$ & $3$ & \multicolumn{1}{l}{$\left( 72\ c2\ Q_{a}-5\ c4\ Q_{b}\right)
/\left( 56\sqrt{30}\right) $} \\ \hline
\end{tabular}
\caption{Same as in table \ref{T8} but for the sub-matrix $^{4}P$ x $^{4}F$.
All the non zero matrix elements are given here, and those corresponding to
the sub-matrix $^{4}F$ x $^{4}P$ are obtained by Hermitian conjugation. The
same abbreviations are used here. }
\label{T9}
\end{table}

\begin{table}[tbp]
\centering
\par
\begin{tabular}{lcccc}
\hline
& g$_{1}$ & g$_{2}$ & g$_{3}$ & g$_{av}$ \\ \hline
&  & ${\rm PO}_{4}$ &  &  \\ \hline
$b_{0}$ & 4.8027 & 4.8027 & 1.9904 & 3.8653 \\
$b_{1}$ & 5.0477 & 4.5577 & 2.2998 & 3.9684 \\
$b_{2}$ & 5.0477 & 4.5577 & 1.9904 & 3.8653 \\ \hline
$c_{0}$ & 5.0435 & 5.0435 & 1.9829 & 4.0233 \\
$c_{1}$ & 5.7379 & 4.3492 & 2.1171 & 4.0680 \\
$c_{2}$ & 5.7379 & 4.3492 & 1.9829 & 4.0233 \\ \hline
&  & ${\rm AsO}_{4}$ &  &  \\ \hline
$b_{0}$ & 4.8723 & 4.8723 & 1.9885 & 3.9110 \\
$b_{1}$ & 5.7637 & 4.2279 & 2.2364 & 4.0760 \\
$b_{2}$ & 5.7637 & 4.2279 & 1.9885 & 3.9934 \\
$b_{3}$ & 5.6402 & 4.1044 & 1.9885 & 3.9110 \\ \hline
$c_{0}$ & 5.0667 & 5.0667 & 1.9818 & 4.0384 \\
$c_{1}$ & 7.1232 & 3.5385 & 1.9854 & 4.2157 \\
$c_{2}$ & 7.1232 & 3.5385 & 1.9818 & 4.2145 \\
$c_{3}$ & 6.8591 & 3.2744 & 1.9818 & 4.0384 \\ \hline
\end{tabular}
\caption{The principal components of the calculated {\bf g} tensor for the
penta-coordinated ${\rm Co}^{2+}$ and their average $g_{av}$ in {\rm Co:Zn}$%
_{2}{\rm (OH)PO}_{4}$ and {\rm Co:Mg}$_{2}{\rm (OH)AsO}_{4}$. The rows $%
b_{0} $ and $c_{0}$ are for the reference trigonal bipyramid, while $b_{1}$,
$b_{2} $, $b_{3}$, $c_{1}$, $c_{2}$, and $c_{3}$ include the effect of
deformations produced by the crystallographically calculated normal modes.
Rows $b_{1}$ and $c_{1}$ include all the normal modes, while $b_{2}$ and $%
c_{2}$ only include the even modes, and in rows $b_{3}$ and $c_{e}$ we have
also put $Q_{1}=Q_{2}=0$. The values of $B$, $D_{s}$, $D_{t}$ and $\protect%
\zeta $ employed here for rows $b_{j}$ ($j=0-3$) are given in row b of table
\ref{T6p}, and those corresponding to rows $c_{j}$ are given in row c of
that table. }
\label{T11}
\end{table}

\begin{table}[tbp]
\centering
\par
\begin{tabular}{lllcc}
\hline
${\bf \Delta H}_{f}$ $(kcal/mol)$ &  & no ion & {\rm Co}$^{2+}$ & {\rm Zn}$%
^{2+}$ \\ \hline\hline
$\left[ {\rm M(OP(OH)}_{3}{\rm )}_{5}\right] ^{2+}$ &  & -$1214.69$ & -$%
2472.73$ & -$868.50$ \\
$\left[ {\rm M(OP(OH)}_{3}{\rm )}_{6}\right] ^{2+}$ &  & -$1440.16$ & -$%
2754.45$ & -$1100.15$ \\ \hline
\end{tabular}
\caption{Heats of formation, in kcal/mol, from PM3 calculations.}
\label{T13}
\end{table}

\begin{figure}[tbp]
\caption[foglio-Fig1]{Diffuse reflectance spectra for the a) {\rm Co}$_{2}$%
{\rm (OH)PO}$_{4}$ and b) {\rm Co}$_{2}{\rm (OH)AsO}_{4}$. The horizontal
scale is linear in the wavelength, but has been labeled employing the
corresponding wavenumbers.}
\label{Fig001}
\end{figure}

\begin{figure}[tbp]
\caption[foglio-Fig2]{ESR spectra of {\rm Co}$^{2+}$ in {\rm Zn}$_{2}${\rm %
(OH)PO}$_{4}$. a) Experimental spectrum. b) Sum of the simulated spectrum
for both the hexa-coordinated and penta-coordinated complexes. c) Simulated
spectrum for the hexa-coordinated complex. The g-values of both the
octahedral and of the penta-coordinated complexes are given in table \ref
{T01}. The arrows in c) show the g-values and their positions for the
octahedral complex, while those in a) correspond to the penta-coordinated
complex. The insert gives the detail of the experimental hyperfine structure
and of the simulated one (around $g_{2}$), attributed to the {\rm Co}$^{2+}$
in the triangular bipyramid.}
\label{Fig01}
\end{figure}

\begin{figure}[tbp]
\caption[foglio-Fig3]{ESR spectra of {\rm Co}$^{2+} $ in {\rm Mg}$_{2}{\rm %
(OH)AsO}_{4}$. Curves a), b), and c), and the meaning of the arrows and
values in curves a) and c) are the same as in figure \ref{Fig01}. }
\label{Fig02}
\end{figure}

\begin{figure}[hbp]
\caption[foglio-Fig4]{Ball-and-stick model of the penta-coordinated metal
clusters with phosphoric acid molecules as ligands. One atom of each type is
labeled in the figure.}
\label{Fig1}
\end{figure}

\begin{figure}[tbh]
\caption[foglio-Fig5]{Hexa-coordinated metal/phosphoric acid cluster. One
atom of each type is labeled in the figure. }
\label{Fig2}
\end{figure}

\begin{figure}[htb]
\caption[foglio-Fig6]{Cobalt/Zinc cluster built from the atomic coordinates
of the {\rm Co}$_{2}${\rm (OH)PO}$_{4}$ unit cell. Part of the phosphate
ions were replaced by phosphoric acid molecules. One atom of each type is
labeled in the figure, which shows the {\rm Co} in the penta-coordinated
position }
\label{Fig3}
\end{figure}


\begin{references}
\bibitem{Riffel}  H. Riffel, F. Zettler and H. Hess. Neus Jahrb. Mineral.
Monatsch. 514 (1975).

\bibitem{Richmond}  W. E Richmond, Am. Mineral., {\bf 25}, 441 (1940).

\bibitem{Hawthorne}  F. C. Hawthorne, Can. Mineral. {\bf 14}, 143 (1976).

\bibitem{Hill}  R. J. Hill, Am. Mineral., {\bf 61}, 979 (1976).

\bibitem{Harrison}  W.T.A., Harrison, J.T. Vaughey, L.L., Dussack, A.J.,
Jacobson, T.E. Martin and G.D. Stucky,{\rm \ J. Solid State Chem.}\ {\bf 114}%
\ 151\ (1995).

\bibitem{Keller}  P. Keller \ Neues. Jahrb. Mineral. Monatsh, 560 (1971)

\bibitem{KellerHZ}  P. Keller, H. Hess and F. Zettler, Neues Jahrb. Miner.
Abh., {\bf 134}, 147 (1979).

\bibitem{Rojo00}  J. M. Rojo, Ph. D thesis, Universidad del Pais Vasco,
Bilbao, 2000

\bibitem{AbragamB}  A. Abragam and B. Bleaney, {\it Electron Paramagnetic
Resonance of Transition Ions} (Clarendon Press, Oxford, 1970).See pag. 449.

\bibitem{GoniFB}  A. Go\~{n}i, L. M. Lezama, T. Rojo., M. E. Foglio, J. A.
Valdivia and G. E. Barberis, {\rm Phys. Rev. B}{\rm \ }{\bf 57 }246{\bf \ }%
(1998).

\bibitem{BeltranP}  F. G. Beltran and F. Palacio, {\rm J. Phys. Chem. }{\bf %
80 }1373\ (1976).

\bibitem{Palacios}  F. Palacio, {\rm J. Phys. Chem. }{\bf 82 }825\ (1978).

\bibitem{JahnT}  H. A. Jahn and E. Teller, {\rm Proc. Roy. Soc.(London)}
{\rm A, }{\bf 161 }220 (1937).

\bibitem{Jahn}  H. A. Jahn, {\rm Proc. Roy. Soc.(London)} {\rm A, }{\bf 164 }%
117 (1937).

\bibitem{RojoMLBR96}  J. M. Rojo, J. L. Mesa, L. Lezama, G. E. Barberis and
T. Rojo, {\rm \ J. Magn. Magn. Mat.}\ {\bf 157/158}\ 493\ (1996).

\bibitem{Rojo1}  T. Rojo, L. Lezama, J.M. Rojo, M. Insausti, M.I. Arriortua
and G. Villeneuve, {\rm Eur. J. Solid State Inorg. Chem.} \ {\bf 29}\ 217\
(1992).

\bibitem{RojoMPLAR97}  M. M. Rojo, J. L. Mesa, J. L Pizarro, L. Lezama, M. I
Arriortua and T Rojo, {\rm \ J. Solid State Chem.}\ {\bf 132}\ 107\ (1997).

\bibitem{Note1}  When we derive the system parameters of the octahedral
complexes from the optical spectra, we have shifted the ground orbital level
by the spin orbit correction, estimated to be $2.5\ \alpha \lambda \sim
-635\ {\rm cm}^{-1}$. This improvement was not used in the penta-coordinated
complex.

\bibitem{VanVleck}  J. H. Van Vleck, {\rm \ J. Chem. Phys.}{\rm \ }{\bf 7
72\ }(1939).

\bibitem{Koster}  G. F. Koster, J. O. Dimmock, R. G. Wheeler and H. Stats,
{\it Properties of the thirty-two point groups} (M.I.T.Press, Cambridge,
Massachusetts, 1963).

\bibitem{Tinkham(1)}  M. Tinkham, {\rm Proc. Roy. Soc.(London) A, }{\bf 236 }%
549{\bf \ }(1956).

\bibitem{PeixotoF}  L. T. Peixoto and M. E. Foglio, {\rm Revista Brasileira
de F\'{i}sica}{\rm \ }{\bf 13 }564{\bf \ }(1983).

\bibitem{Tucker}  E. B. Tucker, {\rm Phys. Rev. }{\bf 143 }264\ (1966).

\bibitem{AbragamP51}  A. Abragam and M. H. L. Pryce, {\rm Proc. Roy.
Soc.(London)} {\rm A, }{\bf 206 }173 (1951).

\bibitem{Pryce}  H. M. L. Pryce, {\rm Proc. Roy. Soc.(London) A, }\ {\bf 63}
25\ (1950).

\bibitem{Wood}  J. S. Wood, {\rm Inorg. Chem. }{\bf 7 }852 (1968).

\bibitem{Griffith}  J. S. Griffith,{\it The theory of transition-metal ions.
}(University Press, Cambridge, England, 1961), see Eq.~(6.37).

\bibitem{FanoR}  U. Fano and G. Racah, {\it Irreducible tensorial sets}%
.(Academic Press, New York, 1959).

\bibitem{NielsenK}  C. W. Nielsen and G. F. Koster,{\it \ Spectroscopic
coefficients for the p}$^{n}${\it , d}$^{n}${\it , and f}$^{n}$ {\it %
configurations. }(MIT Press, Boston, Mass. 1963).

\bibitem{Woodward}  L. A. Woodward, {\it Introduction to the theory of
molecular vibrations and vibration spectroscopy. }(Clarendon Press, Oxford,
1972).

\bibitem{WilsonDC}  E. B. Wilson, J. C. Decius and P. C. Cross, {\it %
Molecular vibrations.} (McGraw-Hill, New York,1955).

\bibitem{pm3}  J.J.P. Stewart,.{\rm \ J. Comput. Chem} {\bf 10}, 209 (1989).

\bibitem{spartan}  SPARTAN package v. 5 (Wavefunction Inc., 1995).

\bibitem{basis}  For basis set definitions, see for instance A. Szabo e N.S.
Ostlund, {\it Modern Quantum Chemistry} (McGraw-Hill, New York, 1989).
\end{references}
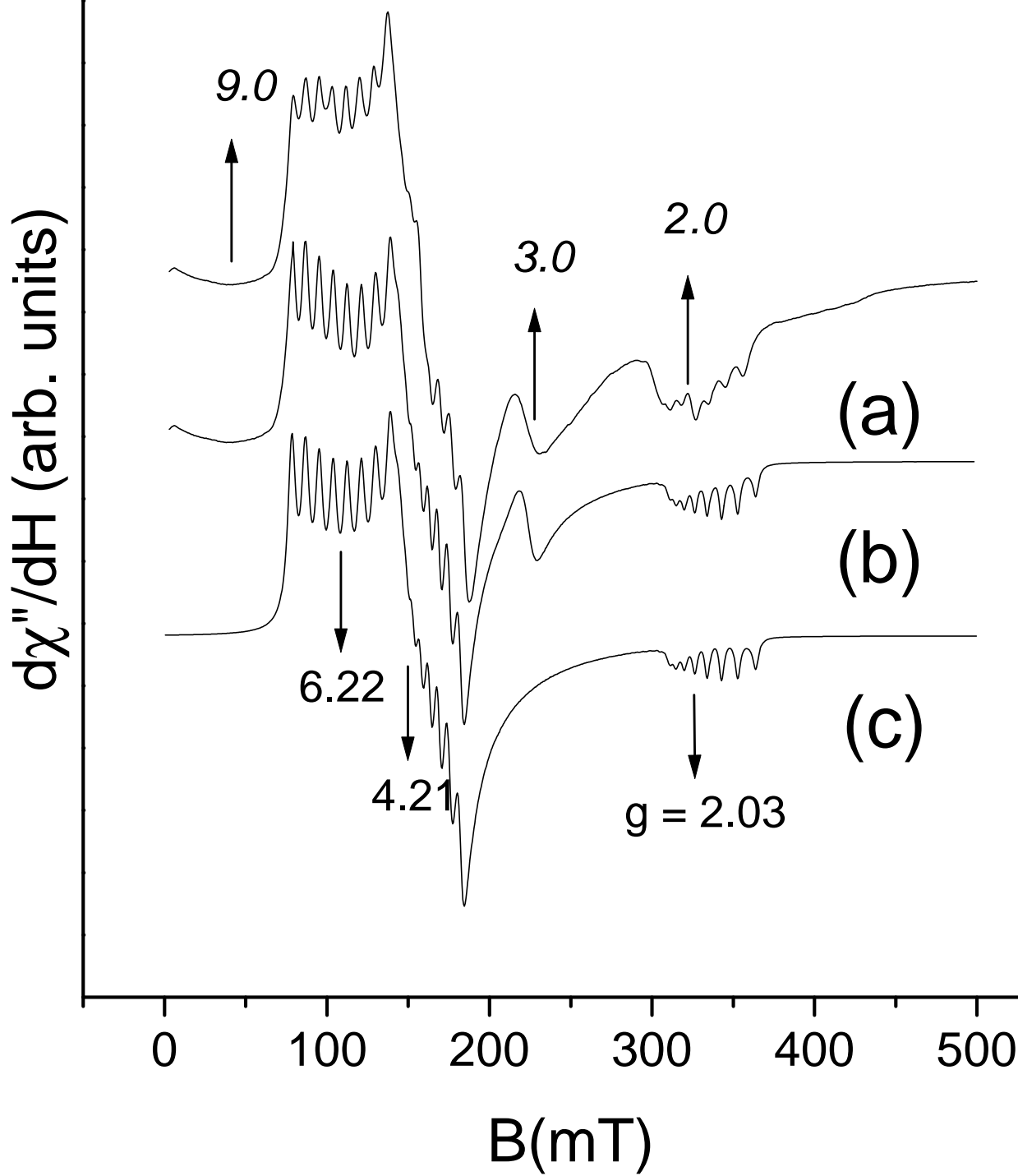
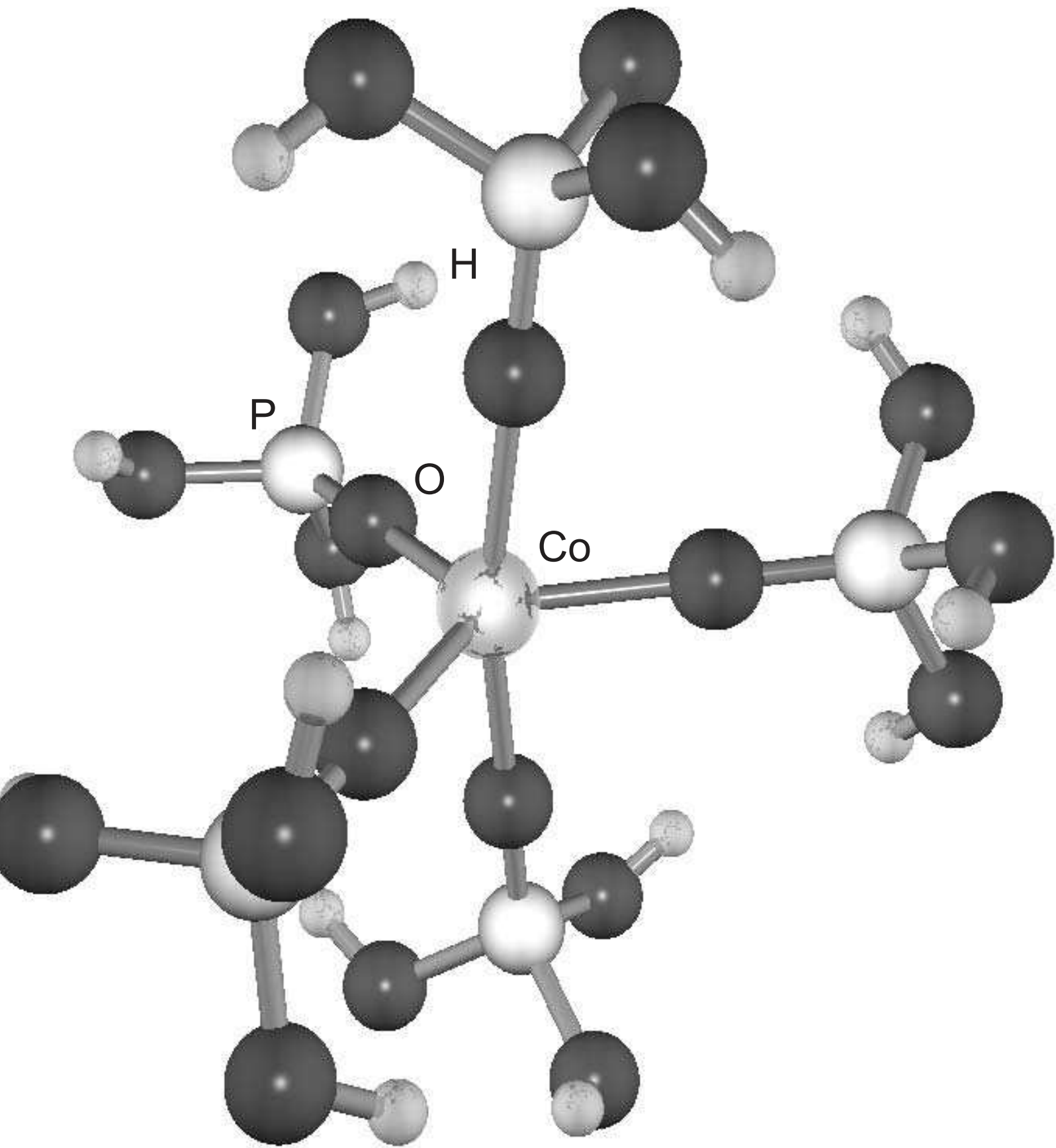
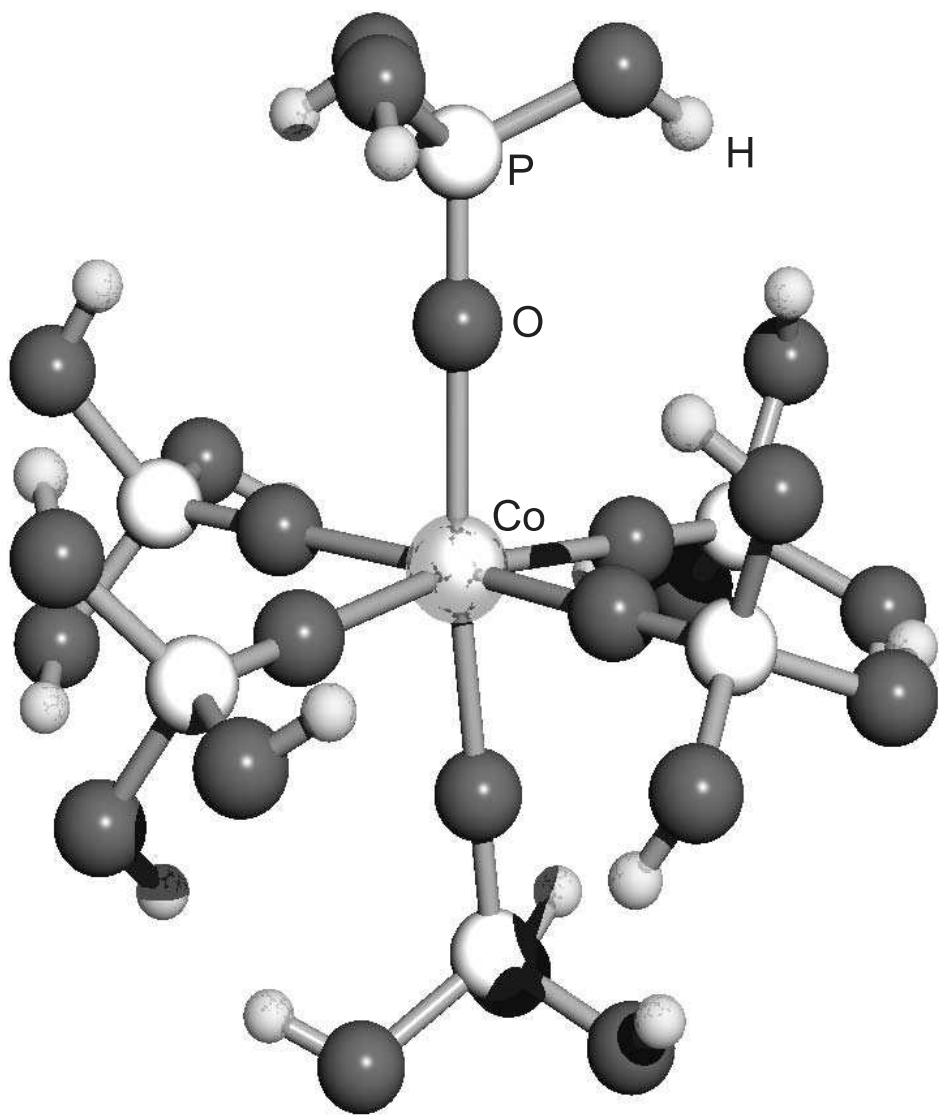
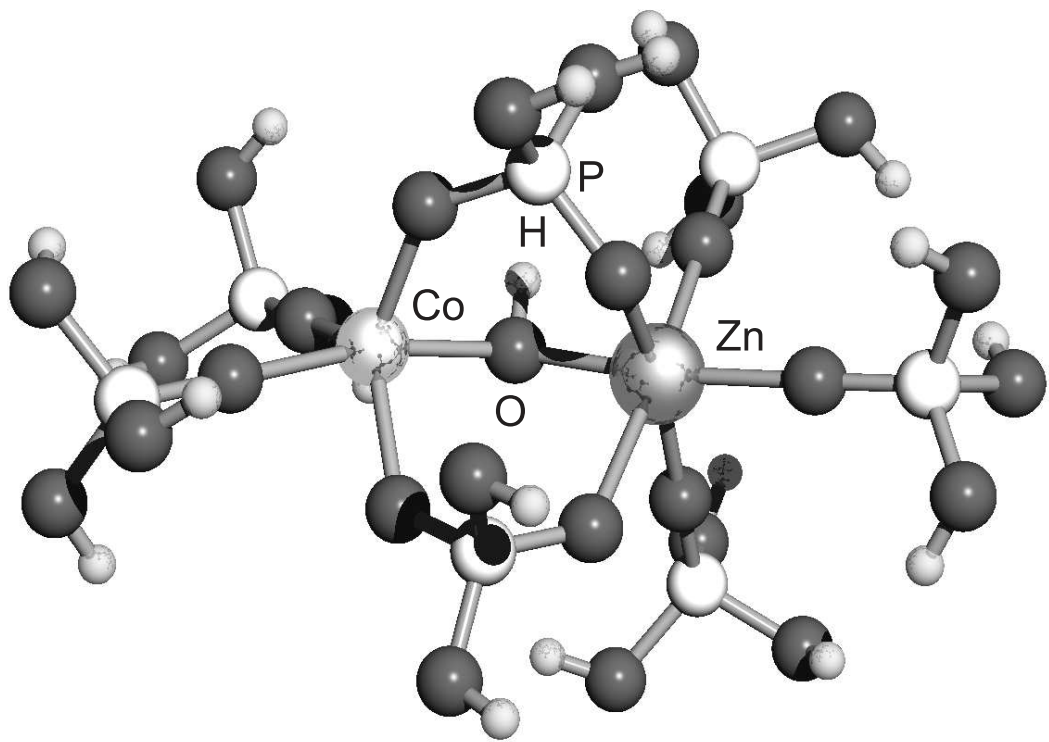
\end{document}